\documentclass[aps,prx,reprint,amssymb,superscriptaddress]{revtex4-2}

\usepackage{xcolor}

\usepackage{amsmath}

\usepackage{bm}

\usepackage{amsmath}  
\usepackage{amsfonts}
\usepackage{braket}
\usepackage{graphicx}
\usepackage{hyperref}
\hypersetup{
    colorlinks=true,
	citecolor=blue,
    linkcolor=red,
	urlcolor = blue
}

\newcommand{\eqfigscl}[2]{\vcenter{\hbox{\includegraphics[scale=#1]{#2}}}}

\begin{document}


\title{Schwinger-Keldysh non-perturbative field theory of open quantum systems beyond the Markovian regime: Application to spin-boson and spin-chain-boson  models}

\author{Felipe Reyes-Osorio}
\affiliation{Department of Physics and Astronomy, University of Delaware, Newark, DE 19716, USA}
\author{Federico Garc\'ia-Gait\'an}
\affiliation{Department of Physics and Astronomy, University of Delaware, Newark, DE 19716, USA}
\author{David J. Strachan}
\affiliation{H. H. Wills Physics Laboratory, University of Bristol, Bristol, BS8 1TL, United Kingdom}
\author{Petr Plech\'a\v{c}}
\affiliation{Department of Mathematical Sciences, University of Delaware, Newark, DE 19716, USA}
\author{Stephen R. Clark}
\affiliation{H. H. Wills Physics Laboratory, University of Bristol, Bristol, BS8 1TL, United Kingdom}
\author{Branislav K. Nikoli\'c}
\email{bnikolic@udel.edu}
\affiliation{Department of Physics and Astronomy, University of Delaware, Newark, DE 19716, USA}


\begin{abstract}

We develop a unified framework for open quantum systems composed of many mutually interacting quantum spins, or any isomorphic systems like qubits and qudits, surrounded by one or more  independent bosonic baths. Our framework, based on Schwinger-Keldysh field theory (SKFT), can handle arbitrary spin value $S$, dimensionality of space, and geometry, while being applicable to a large parameter space for system and bath or their coupling. It can probe regimes in which \textit{non-Markovian dynamics} and nonperturbative effects pose formidable challenges for other state-of-the-art theoretical methods. This is achieved by working with the two-particle irreducible (2PI) effective action, which resums classes of Feynman diagrams of SKFT to an infinite order. Furthermore, such diagrams are generated via an expansion in $1/N$, where $N$ is the number of Schwinger bosons we employ to map spin operators onto canonically commuting ones, rather than via conventional  expansion in system-bath coupling constant. We carefully benchmark our SKFT+2PI-computed results vs. numerically (quasi)exact ones from tensor network calculations applied to the archetypical spin-boson model where both methodologies are applicable. Additionally, we demonstrate the capability of SKFT+2PI to handle a much more complex spin-chain-boson model with multiple baths interacting with each spin where no benchmark from other methods is available at present. The favorable numerical cost of solving integro-differential equations produced by the SKFT+2PI framework with an increasing number of spins and time steps makes it a promising route for simulating driven-dissipative systems in quantum computing, quantum magnonics, and quantum spintronics.
\end{abstract}

\maketitle

\section{Introduction}\label{sec:intro}

The conventional approach to open quantum system dynamics~\cite{Breuer2007,Reimer2019,Breuer2016,Vega2017} formulates  quantum master equations (QMEs) that evolve the reduced density operator of the subsystem of interest via time-local (i.e., differential) equations in the Markovian regime~\cite{Lindblad1976,Nathan2020};  or time-nonlocal~\cite{Nakajima1958,Zwanzig1960} (i.e., integro-differential~\footnote{Although formulating time-local QMEs for non-Markovian dynamics is also possible, this requires handling additional intricacies~\cite{Chruscinski2010,Nestmann2021,Nestmann2021a,Reimer2019}.}) equations in the more general non-Markovian regime~\cite{Rivas2014,Vega2017,Breuer2016}. For many-body systems, the operators within QMEs are usually expressed as polynomials of creation/annihilation operators acting in an exponentially increasing Hilbert space, so that a full matrix representation of the QME quickly becomes intractable by brute force methods. For example, even for a small number $N_S$ of quantum spins of $S=1/2$ or, equivalently, qubits, the size of matrices $2^{N_S} \times 2^{N_S}$ within QME is prohibitively computationally expensive. 

The Markovian limit is characterized by time evolution that depends only on its present state and not on its history. It {\em requires} that  system-environment coupling is weak and that environment correlations are short compared to the timescale of the system evolution (see Ref.~\cite{Nathan2020} and Sec.~\ref{sec:semiclassical} for quantitative criteria involving these parameters to differentiate~\cite{Breuer2016,Vega2017, Rivas2014}  Markovian vs. non-Markovian regimes). For Markovian QMEs where the system contains noninteracting degrees of freedom, so its Hamiltonian is quadratic in creation/annihilation operators, and the dissipative Lindbladian~\cite{Lindblad1976, Nathan2020} operators are linear in creation/annihilation operators, specialized techniques like ``third quantization''~\cite{Prosen2008,Barthel2022, McDonald2023} can dramatically reduce the computational cost---for example, for noninteracting fermions, third quantization replaces $4^{N_e} \times 4^{N_e}$ matrices required in brute force methods with much more manageable $4N_e \times 4N_e$ matrices. Beyond such special cases, the search for polynomially scaling algorithms that can solve many-body Lindblad QME beyond quadratic Hamiltonians has recently led to the development~\cite{Sieberer2016, Thompson2023} of methods based on nonequilibrium quantum field theory in both functional integral~\cite{Rammer2007, Calzetta2008, Berges2015, Gelis2019, Kamenev2023} and second-quantization formulation~\cite{Stefanucci2024}. In particular, the functional integral~\cite{Rammer2007, Gelis2019} formulation of Schwinger-Keldysh field theory 
(SKFT) offers a convenient starting point for calculating expectation values (EVs) and correlation functions of various observables~\cite{Thompson2023,Kamenev2023}, as well as a plethora of field-theoretic tools~\cite{Rammer2007, Gelis2019, Sieberer2016} developed within elementary particle physics. Indeed, SKFT was originally developed for problems in high-energy physics and cosmology~\cite{Calzetta2008, Berges2015, Gelis2019} and later applied to low-energy physics~\cite{Rammer2007, Kamenev2023}. Such SKFT for open quantum systems has been applied to a number of dissipative and/or driven problems~\cite{Maghrebi2016} in condensed matter and atomic-molecular-optical physics. However, while SKFT can, in principle, deal with both Markovian and non-Markovian systems on equal footing, prior works have mostly focused {\em only} on the Markovian regime~\cite{Sieberer2016,Thompson2023,Kamenev2023,Maghrebi2016,McDonald2023,Mueller2017} by building Lindbladian evolution into the functional integral. 

\begin{figure*}
    \centering
    \includegraphics[width=\textwidth]{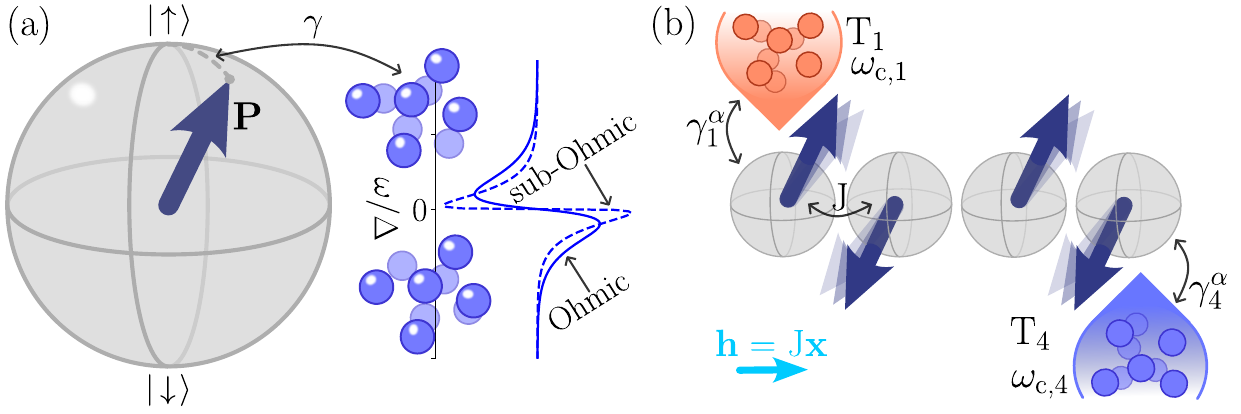}
    \caption{(a) Illustration of the spin-boson model [Eq.~\eqref{eq:sbhamilton}] in which a two-level system, such as spin $S=1/2$ or a qubit, interacts via a system-bath coupling constant $\gamma$ with a dissipative environment modeled as a bosonic bath with infinite frequency content~\cite{Leggett1987}. The bath spectral density is given by Eq.~\eqref{eq:spectral}, where $s=1$ is Ohmic, $0<s<1$ is sub-Ohmic, and $s>1$ is super-Ohmic. The high-frequency content of the bath decays exponentially and is characterized by a cutoff frequency $\omega_c$. The state of the two-level system is uniquely determined by the Bloch vector $\mathbf{P}$, so that $|\mathbf{P}|$ quantifies the purity of the mixed quantum state~\cite{Ballentine2014, Joos2003}. (b) Illustration of the spin-chain-boson model [Eq.~\eqref{eq:chainhamilton}] in which a chain of $N_S=4$ spins $S=1/2$ interact with each other, and can couple to multiple independent baths per spin~\cite{Bruognolo2014,Nemati2022,Hogg2024}. In the chosen system, each component of the spin at the edge of the chain couples to a separate bosonic bath kept at temperatures $T_1$ on the left edge and $T_4$ on the right edge, with $T_1>T_4$ causing thermal transport carried by spin excitations.}
    \label{fig:fig0}
\end{figure*}

On the other hand, many open quantum systems, including notably quantum computers where a dissipative and noisy environment limits operational time of qubits~\cite{Gulacsi2023,Rossini2023}, exhibit pronounced memory effects and thereby time-retarded dynamics. A hallmark of such a non-Markovian regime is the revival of genuine quantum properties, such as quantum coherence, correlations, and entanglement~\cite{Breuer2016,Vega2017}. Such effects are enabled by  backflow of information from the environment to the system. The ability to efficiently simulate non-Markovian open quantum systems also opens new avenues for optimal control~\cite{Koch2016} of open quantum systems, including via environmental engineering~\cite{Harrington2022}. However, the non-Markovian
open quantum systems exhibit exponential growth of complexity with the memory time of the environment, in quite analogous way to how complexity of closed quantum many-body 
systems grows with the number of degrees of freedom. This generally restricts many methods developed for the non-Markovian dynamics to \textit{minimal} system sizes; or, if they can handle larger systems~\cite{Fux2023,Fux2024,Cygorek2024a,Sun2024a, Xu2024,Lindoy2025}, they are typically restricted in evolution times, choice of environment(s)~\cite{Cygorek2024a, Fux2024,Lindoy2025} and spatial geometry~\cite{Fux2023,Sun2024a,Xu2024,Lindoy2025}. 

Thus, these unresolved challenges call for exploration of alternative avenues that could allow one to handle many mutually interacting quantum degrees of freedom, each of which is strongly coupled to possibly \textit{multiple}~\cite{Hogg2024} structured environments. In addition, to describe experimentally relevant systems in quantum computing~\cite{Gulacsi2023,Rossini2023,Rosenberg2017} or quantum spintronics~\cite{Petrovic2021a,Suresh2024,Kovarik2024,Choi2019,Chen2023} and quantum magnonics~\cite{Yuan2022}, one needs to evolve for sufficiently long and experimentally relevant times while handling arbitrary spatial dimensionality (including three-dimensional systems~\cite{Rosenberg2017,Yuan2022}) and geometry~\cite{Patra2025} of both the degrees of freedom and their environments.

In this study, we introduce a field-theoretic approach offering such  capabilities. Our approach combines SKFT with a two-particle irreducible (2PI) effective action~\cite{Rey2004, Berges2015,Borsanyi2005,Babadi2015,Schuckert2018,  Hosseinabadi2023} that  sums~\cite{Brown2015,Mera2016} classes of Feynman diagrams to infinite order. The Feynman diagrams in our SKFT+2PI approach are obtained \textit{not} from conventional perturbative expansion in the system-bath coupling constant, but instead we employ expansion in powers of a nonphysical small parameter $1/N$, where $N$ is the number of Schwinger bosons we employ to map spin operators onto canonically commuting ones. This approach makes it possible to reach regimes that are {\em non-perturbative} in the system-environment coupling constant.  Although both $1/N$ expansion and 2PI resummation techniques were originally developed long ago for problems in high-energy  physics~\cite{Cornwall1974,Gelis2019,Marino2015}, with the understanding that $1/N$ expansion itself  provides a non-perturbative resummation of the conventional perturbation theory, it is only recently~\cite{DiPietro2021} that a more profound  understanding of how the $1/N$ expansion captures non-perturbative effects has been achieved. For example, in many problems, $1/N$ expansion generates~\cite{DiPietro2021}  
terms of a resurgent transseries~\cite{Aniceto2019,Mera2018} in the system-bath coupling constant. 

Our SKFT+2PI approach is developed for a general system of many quantum spins of arbitrary length $S$, their geometry, or their spatial dimensionality, which interact with each other and with one or many environments composed of infinitely many bosonic modes [Fig.~\ref{fig:fig0} and Eq.~\eqref{eq:heishamiltonian}]. Note that  Refs.~\cite{Babadi2015} and~\cite{Schuckert2018} have already employed a similar combination of SKFT, 2PI and $1/N$ expansions to study \textit{closed} systems of quantum spins. Since such closed system dynamics is a limiting case of our more general SKFT+2PI for open quantum systems, in Sec.~\ref{sec:closed} we provide comparison of SKFT+2PI vs. exact results for a small closed system of quantum spins, revealing failure of the former over relatively short time scales. In contrast, our SKFT+2PI for open quantum systems matches surprisingly well numerically (quasi)exact tensor network (TN) methods developed recently. In particular, we carefully benchmark our SKFT+2PI calculations against: the Lindblad QME describing the Markovian dynamics; or methods for non-Markovian dynamics including  hierarchical equations of motion (HEOM)~\cite{Tanimura2020} and TN methods implemented as the time-evolving density operator with orthogonal polynomials algorithm (TEDOPA)~\cite{deVega2015, Chin2010, Ye2021} or  the time-evolving matrix product operator (TEMPO)~\cite{Strathearn2018, Fux2024}. Our results demonstrate that SKFT+2PI can replicate results of these four methods in a large region of the system and bath parameter space. Thus, our SKFT+2PI is a promising \textit{unified framework} for open quantum systems even in the presence of strong interactions and non-Markovian effects. The related recent efforts include Ref.~\cite{Hosseinabadi2023}, where SKFT and 2PI are combined to study open systems of spins $S=1/2$, but only in the Markovian regime.

To demonstrate the capability of our SKFT+2PI approach to probe nonperturbative and non-Markovian features of open quantum systems, we apply it to the \textit{spin-boson} model [Fig.~\ref{fig:fig0}(a)] as an archetypical problem in the field of open quantum systems~\cite{Leggett1987,Breuer2007, Vilkoviskiy2024}. Despite its apparent simplicity [Eq.~\eqref{eq:sbhamilton}], it contains plenty of challenges in the non-Markovian regime that have ignited the development of numerous specialized approaches~\cite{Makri1995,Winter2009,Chin2010,Wang2008e,Wang2010,Strathearn2017,Strathearn2018,Ye2021,Cygorek2022,Weber2022a,deVega2015,Ye2021,Bulla2003,Anders2006,Anders2007,Vojta2012,Orth2013,Vilkoviskiy2024,Scarlatella2024, Ivander2024,Xu2023,Xu2023a}. Furthermore, by applying our SKFT+2PI to a much more complicated \textit{spin-chain-boson} model [Fig.~\ref{fig:fig0}(b)], we demonstrate that it can tackle the previously unsolved  problem of multiple independent environments coupled to noncommuting spin  operators~\cite{Ivander2024, Fux2023,Fux2024,Cygorek2024a,Sun2024a, Xu2024,Lindoy2025}. For both problems, we study the dynamics of spin expectation values, as well as multitime two-spin correlators. In contrast, the knowledge of the reduced density operator obtained from conventional QME approaches~\cite{Shibata1980,Breuer2007,Nakajima1958,Zwanzig1960,Tanimura2020} is \textit{insufficient}~\cite{Fux2023,Fux2024} to obtain multitime correlators.

The paper is organized as follows. The general Hamiltonian on which our SKFT+2PI approach is demonstrated is presented in Sec.~\ref{sec:models}; the spin-boson and spin-chain-boson models are discussed as particular cases of it in Secs.~\ref{sec:spinboson} and~\ref{sec:spinchain}, respectively. Standard methods for solving the spin-boson model are overviewed in Sec.~\ref{sec:sb-methods}, where we also provide details of specific algorithms among those that we employ as benchmarks. In Secs.~\ref{sec:skft} and~\ref{sec:numerical}, the SKFT+2PI theory and its numerical implementation are developed, while Sec.~\ref{sec:initial} discusses how initial conditions are handled. Results obtained with SKFT+2PI for various regimes and specific models are shown in Sec.~\ref{sec:results}. We conclude in Sec.~\ref{sec:conclusions}.

\section{Models}\label{sec:models}

To make the discussion transparent, we focus on a specific and general Hamiltonian
\begin{align}\label{eq:heishamiltonian}
    \hat H &= \mathbf{h}\cdot \sum_{n} \hat{\mathbf{s}}_n + \sum_{nn'} J^{\alpha\beta}_{nn'} \hat s^\alpha_n  \hat s^\beta_{n'} \nonumber \\ 
    &+ \sum_{nk}\omega_{nk} \hat{\mathbf b}^\dagger_{nk} \hat{\mathbf b}_{nk} + \sum_{n\alpha k} g_{nk}^\alpha \hat{s}^\alpha_n \big(\hat{b}^{\alpha\dagger}_{nk} + \hat{b}^{\alpha}_{nk} \big), 
\end{align}
where $\hat{\mathbf{s}}_n = (\hat s^x_n, \hat s^y_n, \hat s^z_n)^T$ are operators for localized quantum spins of length $S$ at sites $n=1\dots N_S$ that are mutually interacting  via the generalized Heisenberg exchange $J^{\alpha\beta}_{nn'}$ and are subject to an external magnetic field $\mathbf{h}$. The system of localized quantum spins $\hat{\mathbf{s}}_n$ is surrounded by an environment composed of many three-dimensional isotropic quantum harmonic oscillators of frequencies $\omega_{nk}$ whose canonical bosonic operators are $\hat{\mathbf b}_{nk} = (\hat b_{nk}^x, \hat b_{nk}^y, \hat b_{nk}^z)^T$. Here, $\hat{\mathbf b}_{nk}$ and $\omega_{nk}$ correspond to the $k$-th oscillator coupled to the $n$-th spin via the in-general anisotropic coupling $g^{\alpha}_{nk}$, Cartesian components are denoted by superscript $\alpha=x,y,z$ and $\hbar=1$ for simplicity of notation. 

\subsection{Spin-boson model}\label{sec:spinboson}

A special case of the general Hamiltonian in Eq.~\eqref{eq:heishamiltonian} describing a single two-level quantum system---such as a spin $S=1/2$, a qubit~\cite{Gulacsi2023,Rossini2023}, or any two well-separated energy levels~\cite{Breuer2016,Vega2017}---which is made open by its interaction with a bosonic bath composed of infinitely many harmonic oscillators is known as the spin-boson model~\cite{Leggett1987}. This model is schematically illustrated in Fig.~\ref{fig:fig0}(a). The Hamiltonian of the spin-boson model is obtained for $N_S=1$ spin of length $S=1/2$ by setting the external magnetic field to $\mathbf{h}=(\Delta,0,\omega_q)$, the Heisenberg exchange to $J_{11}^{\alpha\beta}=0$, and all spin-bath couplings are set to zero except for $g^z_{1k}=g_k$. This leads to simplification of the general Hamiltonian in Eq.~\eqref{eq:heishamiltonian} into
\begin{equation}\label{eq:sbhamilton}
    \hat{H} = \frac{\omega_q}{2} \hat{\sigma}^z + \frac{\Delta}{2} \hat{\sigma}^x + \sum_k \omega_k \hat{b}_k^\dagger \hat{b}_k + \hat{\sigma}^z \sum_k \frac{g_k}{2}(\hat{b}_k + \hat{b}_k^\dagger).
\end{equation}
Here, $\hat{\sigma}^\alpha$ are the Pauli operators; $\omega_q$ is the energy difference between the two eigenstates of $\hat \sigma^z$, \mbox{$\hat{\sigma}^z \ket{\uparrow} =  \ket{\uparrow},\ \hat{\sigma}^z \ket{\downarrow} =  -\ket{\downarrow}$}; $\Delta$ is the tunneling matrix element, which also sets the units of energy; and we suppress the subscript $n$ since there is only one spin. The properties of the bath are fully captured by the coupling-weighted spectral density, $\mathcal J(\omega)=2\pi\sum_k g_k^2 \delta(\omega-\omega_k)$, for which a generic form 
\begin{equation}\label{eq:spectral}
\mathcal J(\omega)=\gamma \omega_c^{1-s} \omega^s e^{-\omega/\omega_c},
\end{equation}
is usually assumed~\cite{Leggett1987}. Here $\gamma$ is a single parameter characterizing the system-bath coupling strength; $\omega_c$ is the cutoff frequency of the bath signifying  the exponential decay of the high-frequency content; parameter $0<s<1$, $s=1$ and $s>1$ classifies spectral densities as sub-Ohmic, Ohmic, and super-Ohmic, respectively~\cite{Wang2010}; and the spectral density $\mathcal J(-\omega)=-\mathcal J(\omega)$ is antisymmetrically extended~\cite{Makri1995}. 

Although the spin-boson model has been intensely studied for many decades~\cite{Leggett1987}, its non-Markovian dynamics~\cite{Rivas2014,deVega2015,Breuer2016,Clos2012,Wenderoth2021} still pose a formidable challenge despite the plethora of available numerical and analytical methods~\cite{Makri1995,Winter2009,Wang2008e,Wang2010,Strathearn2017,Strathearn2018,Ye2021,Cygorek2022,Xu2023,Xu2023a,Weber2022a,Vilkoviskiy2024, deVega2015,Ye2021,Chin2010,Bulla2003,Anders2006,Anders2007,Ivander2024,Vojta2012,Orth2013} for open quantum systems developed specifically for it. This is especially true for the case of zero~\cite{Vilkoviskiy2024,Wang2010,Wenderoth2021} or ultralow temperatures and/or specific frequency content of the bosonic bath~\cite{Clos2012,Wang2010}. For example, the sub-Ohmic case at zero temperature, with a relatively large portion of low-frequency modes [Fig.~\ref{fig:fig0}(a)], is considered particularly challenging~\cite{Wang2010,Xu2023,Bulla2003,Anders2006,Anders2007,Vojta2012} and of relevance to superconducting qubits~\cite{Gulacsi2023,Rossini2023,Rosenberg2017} subjected to electromagnetic noise~\cite{Anders2007}. Other challenges are, in general, posed by the long-time limit~\cite{Wang2010,Xu2022,Xu2023,Xu2023a} of non-Markovian dynamics~\cite{Clos2012,Wenderoth2021}, whose memory effects force information to flow from the environment back into the system. Such effects are not necessarily transient~\cite{Clos2012}. It is generically assumed that the non-Markovian regime~\cite{Rivas2014,Breuer2016,Vega2017} in open quantum system dynamics is entered when the system-bath coupling is sufficiently strong and correlations of the bath do not decay rapidly~\cite{Nathan2020}. But detailed examination~\cite{Clos2012,Wenderoth2021} of measures of non-Markovianity vs. system-bath coupling strength in the case of the spin-boson model shows complex nonmonotonic dependence, including sensitivity to the cutoff frequency $\omega_c$, temperature $T$ and frequency content of the bath.

\subsection{Spin-chain-boson model}\label{sec:spinchain}

Another special case of the general Hamiltonian in Eq.~\eqref{eq:heishamiltonian} is obtained by considering $N_S>1$ spins with Heisenberg exchange coupling between the nearest neighbors $J^{\alpha\alpha}_{n,n+1} = J^{\alpha\alpha}_{n,n-1} = J/2$ and with each component $\alpha=x,y,z$ of the spins at the ends of the chain being coupled to independent bosonic baths [Fig.~\ref{fig:fig0}(b)]. Such coupling to the environment is described by setting all spin-bath couplings to zero except for $g_{1k}^\alpha$ and $g_{N_S k}^\alpha$. An external magnetic field along the $x$-axis is included by using $\mathbf{h} = (h,0,0)$, and coupling between spins is ferromagnetic (FM) if $J < 0$ and antiferromagnetic (AF) if $J > 0$. Thus, the Hamiltonian of such spin-chain-boson model is given by $\hat H = \hat H_S + \hat H_B$, where
\begin{subequations}\label{eq:chainhamilton}
\begin{align}
    \hat H_S &= \sum_{n} \left(J\hat{\mathbf{s}}_n \cdot \hat{\mathbf{s}}_{n+1} + h \hat s^x_n \right), \label{eq:chainhamiltonS} \\ 
    \hat H_B&= \sum_{n=1, N_S}\sum_{k} \left[\omega_{nk} \hat{\mathbf b}^\dagger_{nk} \hat{\mathbf b}_{nk} + \sum_{\alpha} g_{nk}^\alpha \hat s^\alpha_1 (\hat b^{\alpha\dagger}_{nk} + \hat b^\alpha_{nk})\right]. \label{eq:chainhamiltonB}
\end{align}
\end{subequations}
The frequency content of the bosonic baths is  assumed to be described by the $s$-Ohmic form of the spectral density in Eq.~\eqref{eq:spectral}, but with independent temperatures $T_n$, cutoff frequencies $\omega_{c,n}$, Ohmic parameter $s_n$, and system-bath couplings $\gamma_n^\alpha$ for $n=1$ and $n=N_S$. The couplings $\gamma_n^\alpha$ are further assumed to be independent for each spin component $\alpha$, thus allowing for anisotropically coupled baths.  

This spin-chain-boson model represents a challenge to recently developed approaches~\cite{Fux2023,Fux2024,Cygorek2024a,Sun2024a, Xu2024,Lindoy2025, Ivander2024} for non-Markovian dynamics due to its larger Hilbert space and multiple~\cite{Hogg2024} bosonic environments coupled to non-commuting operators. Interestingly, prior studies of open quantum magnets in equilibrium have found that ordinarily forbidden long-range order in one-dimensional (1D) quantum spin chains can be stabilized by the presence of Ohmic and sub-Ohmic baths~\cite{Weber2022}. Thus, having a formalism that can examine non-Markovian dynamics of standard or exotic quantum magnets (such as quantum spin liquids~\cite{Yang2021}), or systems of interacting qubits~\cite{Gulacsi2023,Rossini2023} including their arrangements in three-dimensional geometries~\cite{Rosenberg2017}, is of great contemporary interest, as they are necessarily interacting with a dissipative environment at finite temperature in experiments~\cite{Scheie2021}. While methods based on TNs can be applied to systems similar to the spin-chain-boson model when the coupling to the bath is simplified~\cite{Fux2023, Fux2024}, no standard approach for non-Markovian dynamics in systems with more complicated bath structures (such as the one assumed in this work or with faster entanglement growth occurring in higher spatial dimensions) has emerged thus far.

\section{Standard methods for solving the spin-boson model}\label{sec:sb-methods}

For solving the spin-boson model described in Sec.~\ref{sec:spinboson}, the brute force numerical solution of QME in the non-Markovian regime requires computing high-dimensional integrals over time~\cite{Shibata1980}, where the accuracy of calculations becomes extremely sensitive to numerical errors. As an alternative, the widely used HEOM approach~\cite{Tanimura2020} has been developed by converting the time-nonlocal integro-differential QME of the brute force method~\cite{Shibata1980} into a set of finitely many time-local differential equations. However, its standard version~\cite{Tanimura2020} is limited to high temperatures~\cite{Xu2017}, which has motivated recent efforts to extend the HEOM approach~\cite{Xu2022,Xu2023,Xu2023a} to access zero temperature and much longer simulation times. Multilayer multiconfiguration time-dependent Hartree (ML-MCTDH)~\cite{Wang2008e,Wang2010}, numerical renormalization group (NRG)~\cite{Bulla2003,Anders2006,Anders2007,Vojta2012}, self-consistent dynamical maps~\cite{Scarlatella2024} and TN approaches~\cite{Ivander2024,Fux2024,Cygorek2024a},  that can handle non-Markovian dynamics at zero temperature have also been developed. However, HEOM and ML-MCTDH algorithms are prohibitively expensive for many interacting quantum spins or, equivalently, qubits~\cite{Gulacsi2023,Rossini2023}. 

Since the spin-boson model offers a playground for benchmarking our SKFT+2PI approach vs. standard approaches, we overview in Secs.~\ref{sec:HEOM} and~\ref{sec:tensor} HEOM and specific TN-based algorithms, respectively,  that we employ. We also overview in Sec.~\ref{sec:lindblad} the Lindblad QME we use as a benchmark in the weak system-bath coupling regime.

\subsection{HEOM approach to non-Markovian dynamics}\label{sec:HEOM}

The HEOM algorithm~\cite{Tanimura2020}, initially developed for problems in quantum chemistry~\cite{Tanimura1991}, is a widely used method for solving QMEs of open quantum systems  with arbitrary system-bath coupling. However, in its original formulation~\cite{Tanimura1991, Tanimura2020} it requires finite temperature, \mbox{$T>0$}~\cite{Xu2017,Xu2022,Xu2023,Xu2023a}. The non-perturbative treatment of interaction with the bath is achieved by introducing a hierarchy of auxiliary density matrices that encode system-bath correlations and entanglement~\cite{Tanimura2020, Huang2023}. This hierarchy relies on the expansion of the bath correlation function into an exponential form. The limitations of the HEOM method are well known~\cite{Xu2017,Xu2022,Xu2023,Xu2023a} and arise from the truncation of either the number of auxiliary matrices (a stronger system-bath coupling requires a higher hierarchy cutoff), or the truncation in the exponential decomposition of the bath correlation (typically, lower temperature requires more terms in the expansion). The exact exponential expansion of an arbitrary spectral density $\mathcal J(\omega)$ of the bath is not known. We fit the spectral density in Eq.~\eqref{eq:spectral} using a sum of up to four underdamped~\cite{Tanimura2020} spectral densities whose exponential expansion is well known~\cite{Meier1999}. In order to guarantee convergence, we ran simulations varying the hierarchy cutoff, up to a maximum of 11. In addition, we also adjust the number of exponential terms, using a maximum of 16 terms for the lowest temperature case $k_B T=0.1\Delta$ [Fig.~\ref{fig:heom}], where $k_B$ is the Boltzmann constant. All such calculations were performed using the HEOM extension~\cite{Lambert2023} of the \texttt{QuTiP}~\cite{Johansson2012, Johansson2013} package.

\subsection{Tensor network approach to non-Markovian dynamics}\label{sec:tensor}

The TN approaches to non-Markovian dynamics of open quantum systems are broadly divided into two complementary classes~\cite{Ivander2024}. One class, which hosts TEDOPA used in our study as a benchmark [Figs.~\ref{fig:lindblad},~\ref{fig:heom}, and~\ref{fig:strong}], is based on applying a thermofield chain mapping~\cite{Chin2010,deVega2015}. This approach purifies a finite temperature environment and transforms its representation into a chain geometry ideally suited to matrix product state (MPS) algorithms. Then, the pure quantum state of the full system and environment is unitarily evolved using well established MPS techniques like the time-dependent variational principle (TDVP)~\cite{Haegeman2016,Chanda2020}. The other class is, instead, based on applying a matrix product operator (MPO) to describe the Feynman-Vernon influence functional in the temporal domain~\cite{Strathearn2018,Ye2021,Vilkoviskiy2024,Ng2023,Thoenniss2023,Fux2024}. Separating such representation of the Feynman-Vernon influence functional into contributions from the unitary dynamics and from different additive baths, each of which is represented by a process tensor (PT) leads to increased performance~\cite{Pollock2018,Milz2021,Cygorek2022,Fux2023,Fux2024,Cygorek2024a}. TEMPO, the second TN-based benchmark we employ [Figs.~\ref{fig:lindblad},~\ref{fig:heom}, and~\ref{fig:strong}], belongs to this class. Although TN-based approaches can handle many interacting quantum spins~\cite{Fux2023}, they are also prohibitively expensive in higher dimensions or when time evolution exhibits a transient ``entanglement barrier''~\cite{Lerose2023,Rams2020,Foligno2023}. For example, even Markovian dynamics can lead to a spike~\cite{GarciaGaitan2024} of the many-body entanglement of the system, despite the presence of a dissipative environment and na\"{i}ve expectation~\cite{Trivedi2022} that interactions with the environment should curtail entanglement growth. Moreover, the chain mapping representation of an environment spectral function is also truncated to a finite length, which limits the time that the simulation can reach while faithfully capturing the environment's continuum~\cite{Woods2015,Woods_2016}. The \textit{strongest limitation} of TEDOPA is short time evolution. For TEMPO and PT-TEMPO, there is a combination of Trotter errors and truncation errors from the compression of the MPS bond dimension whose interplay is not currently fully understood and, alas, limits the reachable simulation times~\cite{Ng2023,Thoenniss2023}.

For the spin-boson model [Eq.~\eqref{eq:sbhamilton}] at ultralow temperatures, we use [Figs.~\ref{fig:lindblad}(a),(c) and~\ref{fig:heom}(a),(c)] TEDOPA, based on an MPS description of a thermofield-chain-mapped system~\cite{Chin2010,deVega2015}. An MPS is a representation of an arbitrary pure state as a product of local tensors given by \cite{PhysRevLett.91.147902}
\begin{align}
\ket{\psi} &= \sum_{s_{1},...,s_{N}}A^{s_{1}}_{1}...A^{s_{N-1}}_{N-1}A^{s_{N}}_{N}\ket{s_{1}...s_{N}},
\end{align}
where $A_{j}^{s_{i}}$ is a $\chi_{j}\times\chi_{j+1}$ matrix (with $\chi_{1}=\chi_{N}=1$ fixed) for the $j$th local degree of freedom possessing a $d_j$ dimensional Hilbert space. The bond dimension $\chi_j$ is a crucial parameter controlling the expressiveness of the MPS ansatz and is directly related to the maximum bipartite entanglement it can support. 

\begin{figure}
    \centering
    \includegraphics[width=\columnwidth]{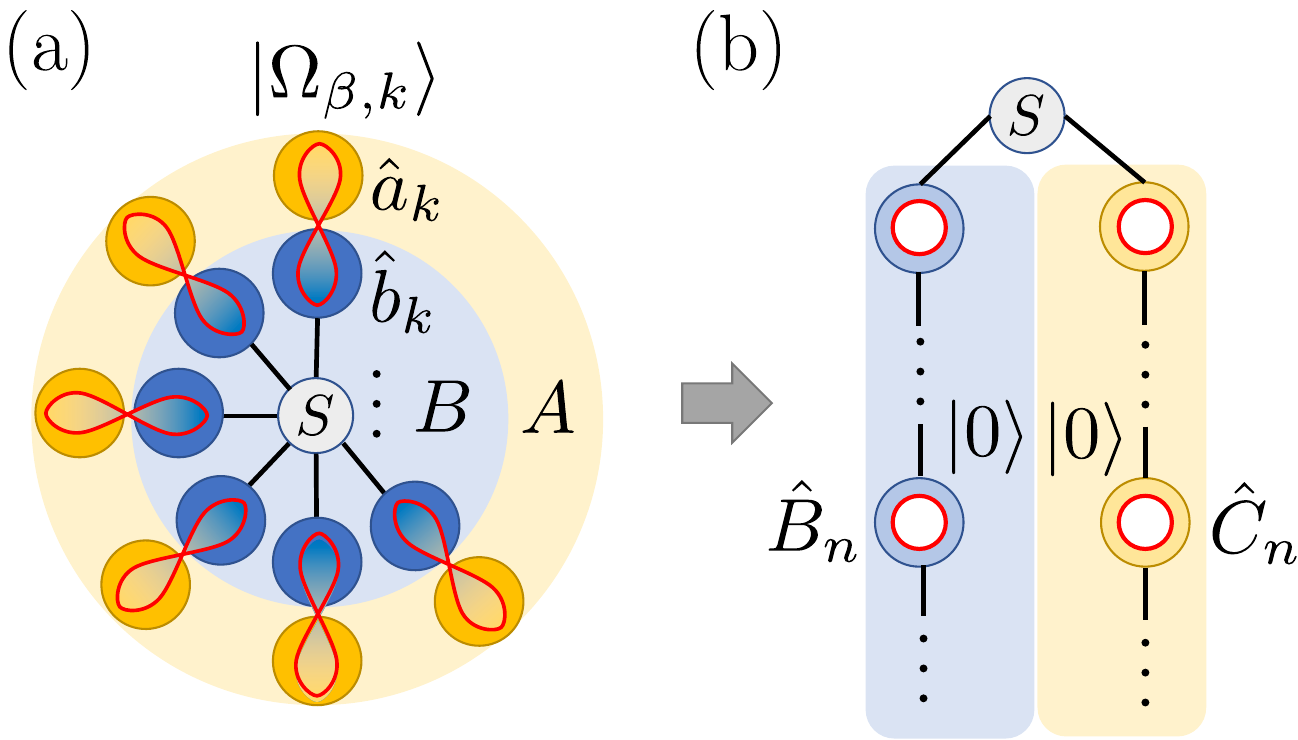}
    \caption{(a) Illustration of how thermofield purification of the initial thermal state of the bath entangles each bath eigenmode with an ancilla. The geometry of interactions of the spin $S$ with the bath modes is a star. (b) After performing a Bogoliubov transformation and orthogonal polynomial tridiagonalization, we arrive at a two-chain geometry ideal for MPS calculations.}
    \label{fig:star_to_chain}
\end{figure}

In order to represent the equilibrium state of the bath in the form of a pure state MPS, we use thermofield purification in which the finite temperature of the bath is encoded in two different baths at zero temperature \cite{takahashi1996thermo}. The density operator of a bosonic bath at inverse temperature $\beta=1/k_B T$ is given by 
\begin{equation}
\hat{\rho}_{\beta} = \bigotimes_{k}\bigg(\sum_{n=0}^{\infty}{\frac{e^{-\beta n\omega_{k}}}{Z_{k}}}\ket{n}_k\bra{n}_{k}\bigg),
\end{equation}
where $Z_{k} = (1-e^{-\beta\omega_{k}})^{-1}$; $n$ is the occupation of mode $k$; and $k_B$ is the Boltzmann constant. By introducing an identical auxiliary system $A$, with canonical operators $\hat{a}^{\dag}_{k},\hat{a}_{k}$ and Hamiltonian $H_{A} = -\sum_{k}\omega_{k}\hat{a}^{\dag}_{k}\hat{a}_{k}$, we define the thermofield double state as a purification of $\hat{\rho}_{\beta}$, given by 
\begin{align}
\ket{\Omega_{\beta}} &= \bigotimes_{k}\bigg(\sum_{n=0}^{\infty}\sqrt{\frac{e^{-\beta n\omega_{k}}}{Z_{k}}}\ket{n}_{k}\otimes\ket{n}^A_{k}\bigg) \nonumber \\
&= \textrm{exp}\bigg(\sum_{k}\theta_{k}(\hat{b}_{k}\hat{a}_{k}-\hat{b}^{\dag}_{k}\hat{a}^{\dag}_{k})\bigg)\ket{\textrm{vac}}.
\end{align}
Here, $\ket{\textrm{vac}}$ is the bosonic vacuum state; $\ket{n}^A_k$ is a state of the $k$ mode in the auxiliary system; \mbox{$\theta_{k} = \textrm{atanh}(e^{-\beta\omega_{k}/2})$}; and $\hat{\rho}_{\beta} = \textrm{Tr}_{A}\ket{\Omega_{\beta}}\bra{\Omega_{\beta}}$ is recovered by partial trace over the states of auxiliary system $A$. The state $\ket{\Omega_{\beta}}$ is the vacuum for the modes 
\begin{align}
\hat{c}_{1,k} &= \textrm{cosh}(\theta_{k})\hat{b}_{k}-\textrm{sinh}(\theta_{k})\hat{a}^{\dag}_{k}, \\
\hat{c}_{2,k} &= \textrm{cosh}(\theta_{k})\hat{a}_{k}-\textrm{sinh}(\theta_{k})\hat{b}^{\dag}_{k},
\end{align}
obtained from a thermal Bogoliubov transformation. In this new basis, the extended Hamiltonian is given by
\begin{align}
    \hat{H} &= \frac{\omega_q}{2} \hat{\sigma}^z + \frac{\Delta}{2} \hat{\sigma}^x + \sum_k \omega_k (\hat{c}_{1,k}^\dagger \hat{c}_{1,k}-\hat{c}_{2,k}^{\dag}\hat{c}_{2,k}) \nonumber \\
    &  + \hat{\sigma}^z \sum_k \left[\frac{g_{1k}}{2}(\hat{c}_{1,k} + \hat{c}_{1,k}^\dagger)+\frac{g_{2k}}{2}(\hat{c}_{2,k} + \hat{c}_{2,k}^\dagger)\right],
\end{align}
where $g_{1k} = g_{k}\textrm{cosh}(\theta_{k})$ and $g_{2k} = g_{k}\textrm{sinh}(\theta_{k})$. 

As it stands, this setup has a star geometry in which the spin interacts with each mode of the bath, as illustrated graphically in Fig.~\ref{fig:star_to_chain}(a). This corresponds to long-ranged interactions within a 1D representation of the system, which are more difficult to handle using MPSs. For this reason, we map via continuous mode tridiagonalization the two zero-temperature star geometry baths into two 1D tight-binding chains, each coupled to the system spin~\cite{deVega2015}, as shown in Fig.~\ref{fig:star_to_chain}(b). In the continuum representation, these two baths are characterized by spectral 
densities $\mathcal J_{1}(k) = [1+n_{\rm BE}(k)]\mathcal J(k)$ and $\mathcal J_{2}(k) = n_{\rm BE}(k)\mathcal J(k)$, where $n_{\rm BE}(k)$ is the Bose-Einstein distribution function. We then define new bosonic operators $\hat B_{n}$ and $\hat C_{n}$ such that 
\begin{alignat}{2}
\hat{c}_{1,k} = \sum_{n}U_{1,n}(k)\hat B_{n},
\quad&
    \quad &
\hat{c}_{2,k} = \sum_{n}U_{2,n}(k)\hat C_{n}.
\end{alignat}
Here, $U_{j,n}(k) = g_{j}(k)\pi_{j,n}(k)/\rho_{n,j}$ for $j=1,2$ and $\pi_{j,n}(k)$ are monic orthogonal polynomials that obey 
\begin{equation}
\int_{0}^{\infty} \! dk \, \mathcal J_{j}(k)\pi_{j,n}(k)\pi_{j,m}(k)=\rho_{j,n}^{2}\delta_{n,m},
\end{equation}
with $\rho_{j,n}^{2} = \int_{0}^{\infty}\! dk \, \mathcal J_{j}(k)\pi_{j,n}^{2}(k)$ \cite{Chin2010}. This description simplifies significantly at zero temperature, as $\mathcal J_{1}(k) = 0$, so only one chain is needed. Using a finite cutoff of $M$ modes for the mapping, we have 
\begin{align}
\hat{H} &= \frac{\omega_q}{2} \hat{\sigma}^z + \frac{\Delta}{2} \hat{\sigma}^x + \hat{\sigma}^{z}\Big[\rho_{1,0}(B_{0}+B_{0}^{\dag}) + \rho_{2,0}(C_{0}+C_{0}^{\dag})\Big] \nonumber \\ 
&\quad  +
\sum_{n=0}^{M}\Big(\alpha_{1,n}\hat{B}^{\dag}_{n}\hat{B}_{n} - \alpha_{2,n}\hat{C}^{\dag}_{n}\hat{C}_{n} + \sqrt{\beta_{1,n+1}}\hat{B}^{\dag}_{n+1}\hat{B}_{n}\nonumber \\
&\quad  - \sqrt{\beta_{2,n+1}}\hat{C}^{\dag}_{n+1}\hat{C}_{n} + {\rm H.c.}\Big),
\end{align}
where the coefficients $\alpha_{j,n}$ and $\beta_{j,n}$ are defined through the recurrence relation $\pi_{j,n+1}(k) = (k-\alpha_{j,n})\pi_{j,n}(k)-\beta_{j,n}\pi_{j,n-1}(k)$, with $\pi_{j,-1}(k) = 0$. These chain parameters were generated using the \texttt{ORTHPOL} package~\cite{Gautschi2005}. Generically, they are found to quickly converge to constants $\alpha_{i,n}\to\alpha_{i}$, $\beta_{i,n}\to\beta_{i}$. Using the Lieb-Robinson bounds \cite{Woods2015,Woods_2016}, sites further than $\sim \tau\beta_{i}$ have a negligible effect on the system dynamics up to time $\tau$, giving a well-defined measure of the length of bath chains we need. In this sense, the discretization generated by orthogonal polynomials is exact up to a finite time. To time evolve the MPS, we use the two-site variant of the TDVP~\cite{PhysRevLett.107.070601,PhysRevB.94.165116,doi:10.1137/140976546,PAECKEL2019167998} which dynamically updates the MPS bond dimensions to maintain a desired level of precision.

\subsection{Lindblad QME approach to Markovian dynamics}\label{sec:lindblad}

In the weak system-bath coupling regime of the spin-boson model [Eq.~\eqref{eq:sbhamilton}], where the system (i.e., spin) dynamics is expected to be Markovian~\cite{Clos2012, Wenderoth2021}, the Lindblad QME~\cite{Lindblad1976, Nathan2020}
\begin{equation}\label{eq:lindblad}
    \frac{d\hat{\rho}}{dt} = -i[\hat{H}_{S}, \hat{\rho}] + \sum_{i=0}^2\hat{L}_i \hat{\rho}\hat{L}_i^\dagger -\frac{1}{2}\{\hat{L}_i^\dagger \hat{L}_i, \hat{\rho}\},
\end{equation}
can accurately capture the open quantum system dynamics. Here $\hat{H}_{S}$ is the Hamiltonian of an isolated spin, composed of the first two terms on the right-hand side (RHS) of Eq.~\eqref{eq:sbhamilton}; $\hat{\rho}$ is the spin density operator~\cite{Ballentine2014}; and $\hat{L}_i$ is a set of three Lindblad operators~\cite{Lindblad1976, Nathan2020} which account for the presence of the bosonic bath. Those three $\hat L_i$ operators for the spin-boson model can be expressed~\cite{Breuer2007} in the energy eigenbasis of $\hat{H}_{S}$, $\hat{H}_{S}\ket{\pm} = E_\pm\ket{\pm}$, of $\hat{H}_\mathrm{S}$ as
\begin{subequations}\label{eq:jumpOps}
    \begin{align}
        \hat{L}_0&=\sqrt{J(\Delta E)[1+n_{\rm BE}(\Delta E)] [\langle +| \hat{\sigma}_z | -\rangle]^2/4 }|-\rangle \langle + |,\\
        \hat{L}_1&=\sqrt{J(\Delta E)n_{\rm BE}(\Delta E) [\langle +| \hat{\sigma}_z | -\rangle]^2/4 }|+\rangle \langle - |,\\
        \hat{L}_2&= \sqrt{\gamma_L T \langle -|\hat{\sigma}_z|-\rangle\langle +|\hat{\sigma}_z|+\rangle/2} |-\rangle \langle -|, 
    \end{align}
\end{subequations}
where $\Delta E=E_+ - E_-$ is the energy difference of the two levels, and $\gamma_L$ is the system-bath coupling. Note that $\gamma_L$ has to be adjusted in Fig.~\ref{fig:lindblad} by hand to match SKFT+2PI- or TN-computed results.


\section{Schwinger-Keldysh field theory + 2PI for open quantum spin systems}\label{sec:skft}

To define the functional integral of SKFT, it is convenient to first map the spin operators in the general Hamiltonian of Eq.~\eqref{eq:heishamiltonian} onto fermionic or bosonic operators~\cite{Bajpai2021, Schuckert2018, Babadi2015} subject to canonical commutation relations, so that the Wick theorem and other field-theoretic machinery applicable to such operators can be utilized. Here we employ the Schwinger boson mapping~\cite{Auerbach1994, Zhang2022, Schuckert2018}, in which operators of spin $S$ are expressed using $N$ flavors of bosons. Any spin length $S$ can be represented by any number of Schwinger boson flavors $N$. Therefore, $N$ is not a physical parameter; rather, it is an auxiliary object of the mathematical framework. For $N=2$, which we use in this work to curtail computational complexity, the spin operators are expressed as 
\begin{equation}
\hat{s}^\alpha_n = \frac{1}{2} \hat\psi^\dagger_n \boldsymbol{\sigma}^\alpha \hat\psi_n,
\end{equation}
where $\boldsymbol{\sigma}^\alpha$ is a matrix representation of the Pauli operators, and $\hat \psi_n=(\hat a_n^{(1)}, \hat a_n^{(2)})^T$ is a doublet of the two flavors of Schwinger bosons for spin $n$. Although other mappings from spin to bosons or fermions can be employed, Schwinger bosons preserve rotational symmetry, as opposed to Holstein-Primakoff bosons~\cite{Gohlke2023}; and are also generalizable to larger spin values $S$, unlike Majorana~\cite{Babadi2015} or Jordan-Wigner~\cite{Li2022} fermions applicable only to $S=1/2$. The spin length $S$ is set by constraining the Schwinger boson occupation at each site to be 
\begin{equation}\label{eq:constraint}
\hat a_n^{(1)\dagger} \hat a_n^{(1)} + \hat a^{(2)\dagger}_n \hat a^{(2)}_n = 2S.
\end{equation}
The constraint ensures that only a subspace of the infinite dimensional bosonic Hilbert space is utilized for spin dynamics, such as $\ket{1,0}\equiv \ket{\uparrow}, \ \ket{0,1} \equiv \ket{\downarrow}$ spanning the physical Hilbert space in the case of spin $S=1/2$.  

The Schwinger-Keldysh (SK) functional integral is formulated in terms of complex fields $\psi_n$ which are complex eigenvalues
\begin{equation}\label{eq:eigen}
\hat\psi_n |\psi_1,\dots\psi_n,\dots \psi_{N_S}\rangle = \psi_n|\psi_1,\dots\psi_n,\dots \psi_{N_S}\rangle, 
\end{equation}
of $\hat\psi_n$, and whose corresponding eigenvectors $|\psi_1,\dots\psi_n,\dots \psi_{N_S}\rangle$ are the bosonic coherent states. Instead of working with complex-valued fields for the Schwinger bosons, we expand in terms of their real and imaginary parts, $a_n^\sigma = (x_{n}^\sigma + i p^\sigma_n)/\sqrt{2}$, which are grouped into the 4-component field 
\begin{equation}\label{eq:fourfields}
\varphi_n = (x_n^{(1)}, p_n^{(1)}, x_n^{(2)}, p_n^{(2)})^T.
\end{equation}
Working with the real-valued fields $\varphi_n$ simplifies the rules that generate the diagrammatic expansion of the 2PI action. Additionally, it manifests the $O(4)$ symmetry of the theory. Since an $O(2)$ theory, which has fewer fields, is already remarkably close to the large-$N$ limit~\cite{Aarts2002}, we conclude that the usage of a $1/N$ expansion~\cite{Marino2015} is justified, despite only employing two flavors of complex Schwinger bosons. 

The spin fields can be constructed as $\sigma^\alpha_n=\varphi^T_n K^\alpha \varphi_n/2$, where 
\begin{equation}\label{eq:realPauli}
    {K}^x = \boldsymbol{\sigma}^x\otimes  I_2, \quad {K}^y = -\boldsymbol{\sigma}^y\otimes \boldsymbol{\sigma}^y, \quad  {K}^z = \boldsymbol{\sigma}^z\otimes  I_2,
\end{equation}
and $ I_2$ is the 2$\times$2 identity matrix. Then, the SK action, $S=S_S+S_B$, as one of the central quantities in SKFT, is given by
\begin{subequations}\label{eq:sbAction}
\begin{align}
    S_S &= -\int_\mathcal{C}\! dt \, \sum_n\varphi^T_n \Big(\frac{i}{2}{K}^0 \partial_t +  H\Big)\varphi_n + \sum_{\alpha\beta nn'} J^{\alpha\beta}_{nn'} \sigma^\alpha_n\sigma^\beta_{n'}, \label{eq:sAction}\\
    S_B &= \int_\mathcal{C}\! dt \, \sum_{n\alpha k} \left[ b^{\alpha \star}_{nk} (i\partial_t-\omega_{nk})b_{nk}^\alpha - g_{nk}^\alpha\sigma_n^\alpha (b_{nk}^\alpha + b^{\alpha\star}_{nk}) \right],  \label{eq:bAction}
\end{align}
\end{subequations}
where $\mathcal{C}$ is the SK closed time contour~\cite{Kamenev2023, Berges2015, Calzetta2008, Gelis2019}. Here $S_S$ and $S_B$ are the contributions to the total action $S$ from the system of spins and the bath, respectively; $ K^0 =  I_2 \otimes \boldsymbol{\sigma}^y$; and $H=\sum_\alpha h^\alpha {K}^\alpha/4$. For simplicity, we consider bosonic baths that are local, i.e., no more than a single spin can couple to a given  bath. This does not preclude multiple baths from coupling to the same spin, which can also be straightforwardly generalized to nonlocal baths. Since $S_B$ is in the Gaussian form, it can be integrated out exactly. This leaves behind a quartic term 
\begin{equation}\label{eq:quartic}
\propto \sigma^\alpha_n(t)\Xi^\alpha_n(t, t^\prime)\sigma^\alpha_n (t^\prime),
\end{equation}
in the total action, representing a \textit{nonlocal-in-time} effective self-interaction of the spins generated by the presence of the bath. The bath kernel 
\begin{equation}\label{eq:bathkernel}
\Xi^\alpha_n(t,t^\prime) = \sum_k (g^\alpha_{nk})^2 B^\alpha_{nk} (t-t^\prime),
\end{equation}
is given in terms of the noninteracting GF of the bosonic bath 
\begin{equation}
iB^\alpha_{nk}(t,t^\prime) = \langle b^\alpha_{nk}(t) b^{\alpha\star}_{nk}(t^\prime) \rangle_0,
\end{equation} 
where $\braket{\dots}_0$ is the nonequilibrium expectation value (EV)~\cite{Kamenev2023} neglecting spin-bath and spin-spin coupling. The posibility of placing the two times $t$ and  $t^\prime$ on the forward and backward branches of the SK contour $\mathcal C$ makes $B^\alpha_{nk}(t,t')$ and any other GF of the theory composite objects consisting of four components~\cite{Gelis2019,Kamenev2023,Stefanucci2013,Schluenzen2019}. However, only two of those components are independent, motivating different expressions in terms of GFs that take real-time arguments. In condensed matter physics, it is common to use either the lesser $G^<$ and greater $G^>$ GFs, often employed in the second-quantized operator formulation of Keldysh GFs~\cite{Stefanucci2013,Schluenzen2019}; or the retarded and Keldysh GFs, often employed in the equivalent functional integral formulation~\cite{Kamenev2023}. In this work, we will decompose contour GFs in terms of their Keldysh ($K$) and spectral ($s$) components~\cite{Berges2015}, respectively expressed as
\begin{equation}
G^{K/s}(t,t') = G^>(t,t') \pm G^<(t,t'),
\end{equation}
where $+$ sign is for $G^K$ and $-$ sign is for $G^s$. The physical meaning of $G^s$ is to describe the density of available states, while $G^K$ describes how those states are occupied. The contour GFs are then reconstructed as
\begin{equation}\label{eq:keldyshgf}
G(t,t') = \frac{1}{2}G^K(t,t') +\frac{1}{2}{\rm sgn}_\mathcal{C}(t,t')G^s(t,t'),
\end{equation}
where the contour sign function ${\rm sgn}_\mathcal{C}(t,t')$ equals 1 if its arguments are ordered on the contour; $-1$ if its arguments are not ordered on the contour; and $0$ if its arguments are the same. Through this decomposition, integrals over $\mathcal C$ simplify to become real-time causal integrals, which, otherwise, in the lesser/greater representation~\cite{Stefanucci2013, Schluenzen2017} require the usage of the much more demanding Langreth rules~\cite{Hyrkaes2019}. For the noninteracting GFs of the bosonic bath $B^\alpha_{nk}(t,t^\prime)$, closed expressions for $B^{\alpha, K/s}_{nk}$ are given by
\begin{subequations}\label{eq:freeBathGF}
\begin{align}
    B^{\alpha, K}_{nk}(t,t') &= -i \coth\left( \frac{\omega_{nk}}{2 k_B T_n} \right) e^{-i \omega_{nk}(t-t')}, \\
    B^{\alpha, s}_{nk}(t,t') &= -i e^{-i \omega_{nk}(t-t')},
\end{align}
\end{subequations}
where $T_n$ is the temperature of the $n$-th bath~\cite{Kamenev2023}. 

The total SK action thus has two quartic terms, the one displayed in Eq.~\eqref{eq:quartic} that stems from integrating out the bosonic environment, and the last term on the RHS of Eq.~\eqref{eq:sAction} that stems from the Heisenberg spin-spin exchange interaction. These quartic terms can both be decoupled through the Hubbard-Stratonovich transformation~\cite{Altland2010}, yielding the modified total action
\begin{align}\label{eq:finalAction}
    S &= \int_\mathcal{C}\! dt \,\bigg[-\sum_n\varphi^T_n\Big(\frac{i}{2} {K}^0 \partial_t +  \tilde H_n \Big)\varphi_n \\
    & + \frac{1}{4} \int_\mathcal{C}\! dt^\prime \sum_{\alpha n}\frac{\lambda^\alpha_n (t) \lambda^\alpha_n(t^\prime)}{\Xi^\alpha_n(t,t^\prime)} + \frac{1}{4} \sum \Lambda^\alpha_n [J^{-1}]^{\alpha\beta}_{nn'} \Lambda^\beta_{n'}\bigg]. \nonumber
\end{align}
Here, the Hubbard-Stratonovich transformation-introduced fields $\lambda^\alpha_n$ and  $\Lambda^\alpha_n$ mediate the nonlocal-in-time spin-bath and the spin-spin interactions, respectively, and $\tilde H_n = \frac{1}{4}\sum_\alpha (h^\alpha +\lambda^\alpha_n + \Lambda^\alpha_n)  K^\alpha$ is the effective Hamiltonian.

\begin{figure}
    \centering
    \includegraphics[width=\columnwidth]{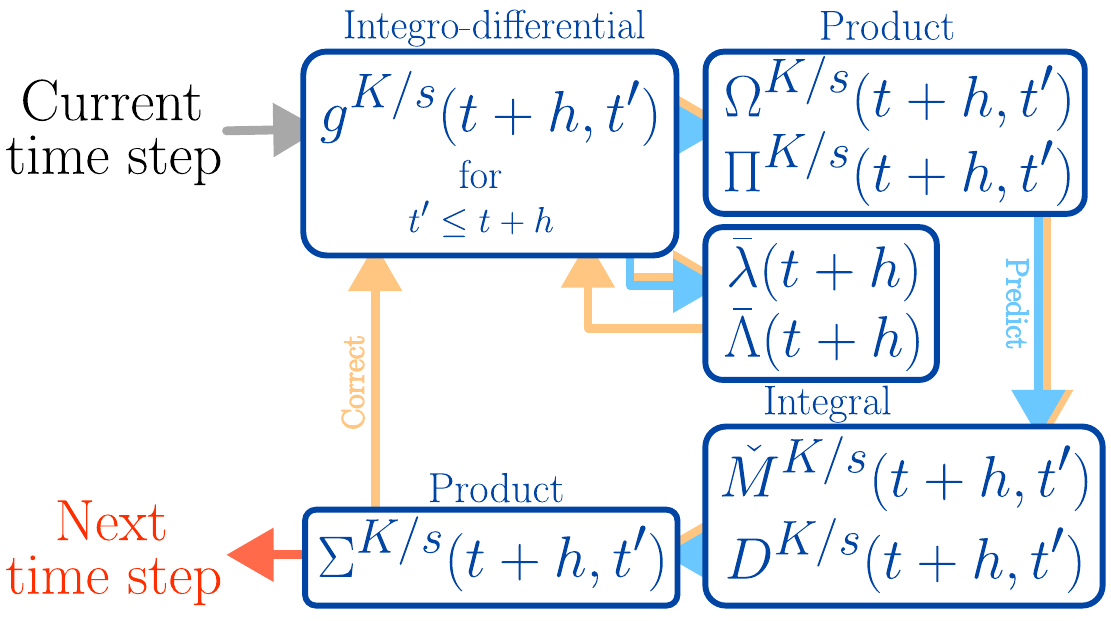}
    \caption{Flowchart of the predictor-corrector algorithm used in the numerical integration of SKFT+2PI-derived coupled Eqs.~\eqref{eq:eomReal}(a)--(n). The type of equation used at each step (integro-differential, integral, or generalized matrix product) is indicated in dark blue. All sub- and super-scripts are omitted for simplicity (see Table~\ref{tab:symbols} for their detailed notation). Functions at time $t$ are used as an input to the prediction stage (light blue arrows), in which all functions are computed  at the next time step $t+h$ for all $t'\leq t+h$, starting with the Keldysh and spectral components of the GFs of the Schwinger bosons $g^{K/s}$. The results of the prediction stage are used to correct $g^{K/s}$, as well as subsequent functions, by repeating the cycle of the self-consistent loop (orange lines). The results of the correction stage are saved as the values for the next time step.}
    \label{fig:flowchart}
\end{figure}

\begin{table*}
\begin{tabular}{lcl | lcl}
Symbol & Definition & Name & Symbol & Definition & Name \\
\hline\hline
$\Xi^\alpha_n(t,t')$          & $\sum_k (g^\alpha_{nk})^2 B^\alpha_{nk} (t-t^\prime)$ & Bath kernel & $\bar\lambda^\alpha_n (t)$    & $\braket{\lambda^\alpha_n(t)}$& EV of the bath field\\ 
$J^{\alpha\beta}_{nn'}$       & & Heisenberg coupling & $\bar\Lambda^\alpha_n (t)$ & $\braket{\Lambda^\alpha_n(t)}$ & EV of the mean-field \\
\hline
$g^{ab}_n(t,t')$              & $-i \langle \varphi^b_n(t')\varphi^a_n(t)\rangle$&  GF of the Schwinger bosons & $\Sigma^{ab}_n(t,t')$         & $2\delta\Gamma_2/\delta g$    & SE of the Schwinger bosons\\
$D^{\alpha}_n(t,t')$          & $-i \langle \lambda^\alpha_n(t')\lambda^\alpha_n(t)\rangle + i\bar\lambda^\alpha_n(t) \bar\lambda^\beta_n(t')$ & Bath propagator & $\Pi^\alpha_n(t,t')$          & $2\delta\Gamma_2/\delta D$    & SE of the bath \\
$M^{\alpha\beta}_{nn'}(t,t')$ & $-i \langle \Lambda^\alpha_{n'}(t')\Lambda^\beta_n(t)\rangle + i\bar\Lambda^\alpha_n(t) \bar\Lambda^\beta_n(t')$  & Mean-field propagator & $\Omega^{\alpha\beta}_n$      & $2\delta\Gamma_2/\delta M$    & SE of the mean-field \\
$\check M^{\alpha\beta}_{nn'}(t,t')$ & Eq.~\eqref{eq:mflocal} & Mean-field propagator & & & \\
\hline
$\tilde H^{ab}_n$             & $ \frac{1}{4}\sum_\alpha (h^\alpha + \lambda^\alpha_n + \Lambda^\alpha_n) K^\alpha_{ab}$ & Effective Hamiltonian & $K^\alpha_{ab}$               & Eq.~\eqref{eq:realPauli} & Spin matrices for real fields   \\
\end{tabular}
\caption{\label{tab:symbols} Summary of the key quantities of the SKFT+2PI framework for open quantum spin systems that appear in the equations of motion, Eqs.~\eqref{eq:eomContour} and~\eqref{eq:eomReal}, with their respective symbols, definitions, and labels. All indices are shown explicitly, such as $a,b=1\dots 4$, $\alpha,\beta=x,y,z$, and $n=1\dots N_S$. }
\end{table*}

Nonequilibrium connected EVs can be obtained from the functional derivatives of the generating functional
\begin{align} \label{eq:wfunc}  
W[J, K] &= -i \ln \int\! \mathcal{D}\Phi \, \exp\bigg(iS[\Phi] + i\int_\mathcal{C}\! dt \, J(t)\Phi(t) \nonumber \\
&+ \int_\mathcal{C}\! dtdt^\prime \, K(t,t^\prime)\Phi(t)\Phi(t^\prime)\bigg),
\end{align}
where $\mathcal D \Phi$ indicates functional integration over all possible configurations of six-component field $\Phi=(\varphi, \lambda, \Lambda)^T$, and $J$ and $K$ are one- and two-particle sources~\cite{Gelis2019, Schuckert2018}, respectively. The Legendre transform of the functional $W[J, K]$ with respect to both arguments is the 2PI effective action~\cite{Berges2015, Rammer2007, Gelis2019} 
\begin{align}\label{eq:gammafunc}
    \Gamma[\bar \Phi, G] &= W[J, K] - \int_\mathcal{C}\! dt\, J(t)\bar \Phi(t) \\
    &- \int_\mathcal{C}\! dtdt^\prime\, K(t,t^\prime)\Big(G(t^\prime,t)-i\bar\Phi(t^\prime)\bar\Phi(t)\Big). \nonumber
\end{align} 
Here, $\bar \Phi$ and $G$ are the one- and two-particle connected EVs generated by $W[J,K]$. That is, $\bar \Phi$ is the EV and $G$ is the connected GF of the fields. It is advantageous to work with $\Gamma[\bar\Phi,G]$ instead of $W[J, K]$ because it produces EVs via a comparatively simpler variational approach, i.e., the full nonequilibrium EVs satisfy the saddle-point equations $\delta \Gamma/\delta\bar\Phi=0$ and $\delta \Gamma/\delta G=0$. Such variational calculations for real fields can be performed on the expansion~\cite{Cornwall1974}
\begin{equation}\label{eq:2piAction}
    \Gamma[\bar\Phi, G] = S[\bar\Phi] + \frac{i}{2}{\rm Tr}\ln G^{-1} + \frac{i}{2}{\rm Tr}\left[G^{-1}_0[\bar\Phi]G \right] - i\Gamma_2,
\end{equation}
where a constant term has been ignored; the trace is taken over all possible indices and times;  \mbox{$G_0^{-1} = \delta^2 S/\delta\bar\Phi\delta\bar\Phi$} is the inverse of the noninteracting GF including one-loop or mean-field corrections; and $\Gamma_2$ contains all 2PI vacuum Feynman diagrams. The 2PI diagrams contain two or more loops in which edges represent the full nonequilibrium GF $G$ and vertices correspond to interactions contained in the action $S$, with possible insertions of the EV of the field $\bar\Phi$. Let us recall that 2PI diagrams are those that cannot be separated by cutting two edges or fewer, and vacuum diagrams have no external edges~\cite{Gelis2019}. Since the diagrammatic expansion contained in $\Gamma_2$ is a functional of the \textit{full interacting} nonequilibrium GF, each diagram within the 2PI effective action formalism effectively sums an infinite number of Feynman diagrams of particular topology with bare edges~\cite{Brown2015}. This can unravel effects that are \textit{non-perturbative}~\cite{Marino2015, DiPietro2021}, which would otherwise be unattainable when using standard perturbative expansion in the coupling constant~\cite{Takei2019, Mahfouzi2014, Schluenzen2019}.

The spin-to-Schwinger-boson mapping implies that any EV containing an odd number of Schwinger bosons vanishes for physical states. In particular, \mbox{$\bar \varphi = 0$} and \mbox{$\langle \varphi(t)\lambda(t^\prime)\rangle = \braket{\varphi(t)\lambda(t^\prime)} = 0$}. Additionally, \mbox{$\braket{\lambda(t) \Lambda(t')} = 0$} due to the absence of terms coupling $\lambda$ and $\Lambda$ in the action of Eq.~\eqref{eq:finalAction}. Therefore, each diagram in the 2PI expansion is made up of vertices
\begin{equation}\label{eq:vertex}
    \eqfigscl{0.25}{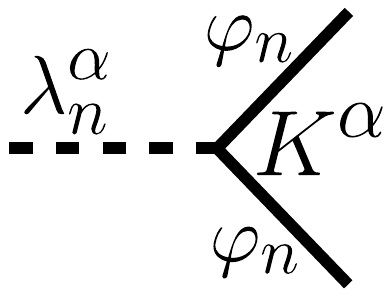}, \quad \text{and} \quad
    \eqfigscl{0.25}{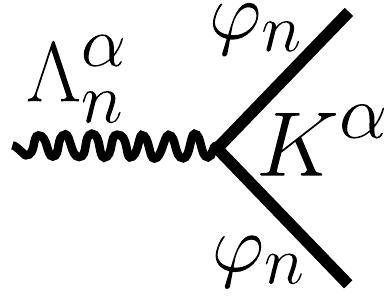},
\end{equation}
where, within a particular diagram, solid lines correspond to the GF of the Schwinger bosons \mbox{$i g^{ab}_{nn'}(t,t')= \braket{\varphi^a_n(t) \varphi^b_{n'}(t')}$}; dashed lines to the propagator \mbox{$i D^\alpha_n = \braket{\lambda^\alpha_n(t) \lambda^\alpha_n(t')}$}; and wavy lines to the propagator \mbox{$iM^{\alpha\beta}_{nn'}=\braket{\Lambda^\alpha_n(t) \Lambda^\beta_{n'}(t')}$}.
Table~\ref{tab:symbols} summarizes our nomenclature for all non-vanishing EVs and other relevant quantities.

Handling the infinite number of diagrams contained within $\Gamma_2$ in Eq.~\eqref{eq:2piAction} requires a {\em controlled approximation} scheme. We adopt the scheme used in Ref.~\cite{Schuckert2018}, where diagrams are truncated based on powers of the inverse of the number of Schwinger bosons $1/N$, which has been previously used to capture relevant features of closed quantum systems~\cite{Berges2002, Aarts2002, Rey2004, Schuckert2018, Babadi2015, Burchards2022, Kronenwett2011}. We emphasize that $N=2$ flavors of Schwinger bosons, which translate into an $O(4)$ theory, have been shown to already be remarkably close to the large-$N$ limit in spite of our $N=2$ not being as large as typically  invoked in elementary particle physics~\cite{Marino2015, Aarts2002}. The scaling with $1/N$ of a particular diagram is set by the number of closed loops of solid lines, and the number of dashed and wavy lines. Closed loops of solid lines scale $\sim N$ due to corresponding to traces over the space of Schwinger bosons. To determine the scaling of dashed (wavy) lines, we consider that all terms in the action of Eq.~\eqref{eq:finalAction} are relevant in the large-$N$ limit on the proviso that the bath kernel and the Heisenberg coupling constant  scale as $\Xi \sim J \sim 1/N$ because the fields $\lambda$ and $\Lambda$ do not scale with $N$ themselves. Then, the equations of motion Eqs.~\eqref{eq:dContour} and~\eqref{eq:mContour}, in which $D\sim \Xi$ and $M \sim J$, imply that dashed (wavy) lines scale $\sim 1/N$. Thus, the 2PI diagrams that scale with the lowest power of $1/N$ are the two-loop ones:
\begin{align}\label{eq:gamma2}
    \Gamma_2 &= \eqfigscl{0.25}{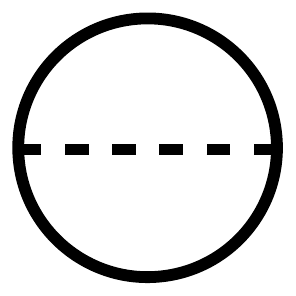} + \eqfigscl{0.25}{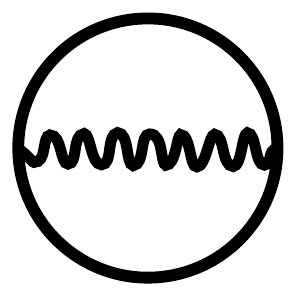} \sim N^0.
\end{align}
At this order of truncation, the equations of motion for the EVs of the fields and the connected nonequilibrium GFs, as obtained from the expansion of the 2PI action in Eq.~\eqref{eq:2piAction} via variational principle, are given by
\begin{widetext}
\begin{subequations}\label{eq:eomContour}
\begin{align}
    \partial_t g_n(t,t') &= i K^0\delta_{tt'}+ 2i K^0\tilde H_n(t) g_n(t,t') + i K ^0\int_\mathcal{C}\! dt_1\, \Sigma_n(t,t_1) g_n(t_1,t'), \\
    M(t,t') &=  2J\delta_{tt'} + 2\int_\mathcal{C}\! dt_1 \, J\Omega(t,t_1) M(t_1, t'), \label{eq:mContour}\\
    D^\alpha_n(t,t') &= 2\Xi^\alpha_n(t,t')
    +  2\iint_\mathcal{C}\! dt_1dt_2 \,\Xi^\alpha_n(t,t_1)\Pi^\alpha_n(t_1,t_2) D^\alpha_n(t_2,t'), \label{eq:dContour}\\
    \bar\Lambda(t) &= \frac{i}{2} J \, {\rm Tr_S}\left[K g (t,t) \right], \label{eq:LambdaContour}\\
    \bar\lambda^\alpha_n(t) &= \frac{i}{2} \int_\mathcal{C}\! dt' \, \Xi^\alpha_n(t,t') {\rm Tr_S}\left[K^\alpha g_{n}(t',t')\right].
\end{align}
\end{subequations}
\end{widetext}
We remind the reader that definitions for all quantities in these equations are provided in Table~\ref{tab:symbols}. Here, matrix multiplication is assumed for indices not shown; ${\rm Tr_S}$ traces over the space of Schwinger bosons; $\delta_{tt'}$ is the contour Dirac delta function~\cite{Berges2015}; integrals are over the SK closed contour $\mathcal{C}$; and \mbox{$\Pi = 2\delta\Gamma_2/\delta D$}, \mbox{$\Omega = 2\delta\Gamma_2/\delta M$}, and \mbox{$\Sigma= 2\delta\Gamma_2/\delta g$} are the self-energies (SEs) derived through functional differentiation of the 2PI vacuum diagrams. The diagrams considered within our $1/N$ expansion [Eq.~\eqref{eq:gamma2}] yield expressions for the SEs
\begin{subequations}\label{eq:selfContour}
    \begin{align}
    \Pi_n^\alpha(t,t') &= \frac{i}{8}{\rm Tr_S}\left[K^\alpha g_n(t,t')K^\alpha g_n^T(t,t')\right], \\
    \Omega^{\alpha\beta}_n(t,t') &= \frac{i}{8}{\rm Tr_S}\left[K^\alpha g_n(t,t')K^\beta g_n^T(t,t')\right], \\
    \Sigma^{ab}(t,t') &=  \frac{i}{4}\sum_\alpha K^\alpha g_n(t,t') K^\alpha D^\alpha_n(t,t') \nonumber \\ 
    & + \frac{i}{4}\sum_{\alpha\beta} K^\alpha g_n(t,t') K^\beta M^{\alpha\beta}_{nn}(t,t'). 
\end{align}
\end{subequations}
Although the GF of the Schwinger bosons $g_{nn'}$ can in principle be nonlocal in space, such as in spin systems with fractional excitations~\cite{Zhang2022, Savary2017}, the separable initial spin states we consider in this work imply $g_{nn'}=g_n \delta_{nn'}$ at all times, and, therefore $\Sigma_{nn'}, \Omega_{nn'}, \Pi_{nn'} \propto \delta_{nn'}$. The self-consistency~\cite{Brown2015} built into 2PI resummation and the SEs evades~\cite{Borsanyi2005} the so-called secularity problem for expansion in terms of the free Keldysh GFs, where elapsed time appearing next to the coupling constant makes the effective coupling arbitrarily large at late times. The same self-consistency ensures that all conservation laws are satisfied in spite of truncating the diagrammatic expansion~\cite{Stefanucci2013}. 

Equations~\eqref{eq:eomContour} and~\eqref{eq:selfContour}, which utilize integrals over the SK contour $\mathcal{C}$, can be transformed to contain real-time integrals by decomposing all GFs and SEs into their Keldysh and spectral components according to Eq.~\eqref{eq:keldyshgf}. In addition to the $M^K$ and $M^s$ components, the propagator $M(t,t')$ has a time-local part proportional to $\delta_{tt'}$ [Eq.~\eqref{eq:mContour}]. The time-local part can be extracted by defining $\check M(t,t')$ such that 
\begin{equation}\label{eq:mflocal}
    M(t,t') = 2 J\delta_{tt'} + J \check{M}(t,t')J.
\end{equation}
Assuming that there is no single-ion anisotropy, i.e., $J^{\alpha\beta}_{nn}=0$, no other GFs or SEs have a time-local component. Therefore, the real-time equations of motion are given by
\begin{widetext}
\begin{subequations}\label{eq:eomReal}
\begin{align}
\partial_t g^K_n(t,t') &= 2iK^0\tilde H_n(t) g^K_n(t,t') + iK^0\int_0^t\!dt_1\, \Sigma_n^s (t,t_1)g^K_n(t_1,t') - iK^0\int_0^{t'}\! dt_1\, \Sigma^K_n (t,t_1)g^s_n(t_1,t') , \label{eq:eomRealF}\\
\partial_t g^s_n(t,t') &= 2iK^0\tilde H_n(t) g^s_n(t,t') + iK^0\int_{t'}^t \! dt_1\, \Sigma^s_n (t,t_1)g^s_n(t_1,t'), \label{eq:eomRealRho}\\
\check M^{K}(t,t') &= 4\Omega^K(t,t') + 2\int_0^t\!dt_1\, \Omega^s (t,t_1) J \check M^K(t_1,t') - 2\int_0^{t'}\! dt_1\, \Omega^K(t,t_1) J \check M^s(t_1,t') \\
\check M^{s}(t,t') &= 4\Omega^s(t,t') + 2\int_{t'}^{t}\!dt_1\, \Omega^s (t,t_1) J \check M^s(t_1,t')  \\
D^{\alpha K}_n(t,t') &= 2\Xi^{\alpha K}_n(t,t') + 2\int_0^t \! \int_0^{t_1} \! dt_1 dt_2 \, \Xi^{\alpha s}_n(t,t_1)\Pi^{\alpha s}_n (t_1,t_2) D^{\alpha K}_n(t_2,t') \\
\displaybreak
&+ 2\int_0^{t'} \! \int_0^{t_1} \! dt_1dt_2 \,\Xi^{\alpha K}_n(t,t_2) \Pi^{\alpha s}_n (t_2,t_1) D^{\alpha s}_n(t_1,t') - 2\int_0^t \! \int_0^{t'} \! dt_1 dt_2 \,\Xi^{\alpha s}_n(t,t_1) \Pi^{\alpha K}_n (t_1,t_2) D^{\alpha s}_n(t_2,t'),\nonumber \\
D^{\alpha s}_n(t,t') &= 2\Xi^{\alpha s}_n(t,t') + 2\int_{t'}^t \! \int_{t'}^{t_1} \! dt_1dt_2 \,\Xi^{\alpha s}_n(t,t_1) \Pi^{\alpha s}_n (t_1,t_2) D^{\alpha s}_n(t_2,t'), \\
\Sigma^K_n(t,t') &= \frac{i}{8}\sum_\alpha \left\{K^\alpha g^K_n(t,t') K^\alpha D^{\alpha K}_n(t,t') + K^\alpha g^s_n(t,t') K^\alpha D^{\alpha s}_n(t,t') \right\}  \\
&+ \frac{i}{8}\sum_{\alpha\beta} \left\{K^\alpha g^K_n(t,t') K^\beta [J\check M^K(t,t') J]^{\alpha\beta}_{nn} + K^\alpha g^s_n(t,t') K^\beta [J\check M^s(t,t') J]^{\alpha\beta}_{nn} \right\}, \nonumber\\
\Sigma^s_n(t,t') &= \frac{i}{8}\sum_\alpha \left\{K^\alpha g^K_n(t,t') K^\alpha D^{\alpha s}_n(t,t') + K^\alpha g^s_n(t,t') K^\alpha D^{\alpha K}_n(t,t') \right\} \\
&+ \frac{i}{8}\sum_{\alpha\beta} \left\{K^\alpha g^K_n(t,t') K^\beta [J\check M^s(t,t') J]^{\alpha\beta}_{nn} + K^\alpha g^s_n(t,t') K^\beta [J\check M^K(t,t') J]^{\alpha\beta}_{nn} \right\}, \nonumber\\
\Omega^{\alpha\beta K}_n(t,t') &= \frac{i}{16}{\rm Tr_S}\left[K^\alpha g^K_n(t,t') K^\beta g^{K \, T}_n(t,t') + K^\alpha g^s_n(t,t') K^\beta g^{s \, T}_n(t,t')\right], \label{eq:eomRealOmegaf}\\
\Omega^{\alpha\beta s}_n(t,t') &= \frac{i}{8}{\rm Tr_S}\left[K^\alpha g^K_n(t,t') K^\beta g^{s \, T}_n(t,t')\right], \label{eq:eomRealOmegarho}\\
\Pi^{\alpha K}_n(t,t') &= \frac{i}{16}{\rm Tr_S}\left[K^\alpha g^K_n(t,t') K^\alpha g^{K \, T}_n(t,t') + K^\alpha g^s_n(t,t') K^\alpha g^{s \, T}_n(t,t')\right], \label{eq:eomRealPif}\\
\Pi^{\alpha s}_n(t,t') &= \frac{i}{8}{\rm Tr_S}\left[K^\alpha g^K_n(t,t') K^\alpha g^{s \, T}_n(t,t')\right], \label{eq:eomRealPirho}\\
\bar\Lambda^\alpha_n(t) &= \frac{i}{4}\sum_{\beta n'} J^{\alpha\beta}_{nn'} {\rm Tr_S}\left[K^\beta g^K_{n'}(t, t)\right], \\
\bar\lambda^\alpha_n(t) &= \frac{i}{4}\int_0^t \! dt_1 \, \Xi^{\alpha s}_n (t,t_1) {\rm Tr_S}\left[K^\alpha g^K_n(t_1, t_1)\right].
\end{align}
\end{subequations}
\end{widetext}

\subsection{Spin expectation values and two-spin correlators from SKFT+2PI}

The dynamics of spin EVs is obtained from the SKFT+2PI equations of motion from
\begin{equation}\label{eq:spinEV}
    \langle \hat s^\alpha_n \rangle (t) = \frac{i}{8}{\rm Tr_S}\left[g^K_n(t,t) K^\alpha\right].
\end{equation}
As such, the EV of the Hubbard-Stratonovich fields can be expressed as 
\begin{subequations}\label{eq:fields}
\begin{align}
    \bar \Lambda^\alpha_n(t) &= 2\sum_{\beta n'} J^{\alpha\beta}_{nn'} \langle \hat s^\alpha_n \rangle (t), \label{eq:meanfield}\\
    \bar \lambda^\alpha_n(t) &= \int_0^t\! dt_1 \, \Xi^{\alpha s}_n(t,t_1) \langle \hat s^\alpha_n \rangle (t_1). \label{eq:pastfield}
\end{align}
\end{subequations}
The RHS of Eq.~\eqref{eq:meanfield} is proportional to the magnetic mean field that spin $n$ is subject to due to being coupled to other spins via Heisenberg exchange. The dynamics of the propagator of the mean field, $iM^{\alpha\beta}_{nn'}(t,t')=\braket{\Lambda^\alpha_n(t) \Lambda^\beta_{n'}(t')}$, includes all fluctuations of $\Lambda^\alpha_n$. Similarily, the RHS of Eq.~\eqref{eq:pastfield} is an average over all past spin states weighed by the spectral component of the bath kernel $\Xi^{\alpha s}_n$, highlighting the generally non-Markovian nature of the dynamics.

The two-spin connected correlator functions $\braket{s^\alpha_n(t) s^\beta_{n'}(t')}$ in the SKFT+2PI approach are obtained from the connected generating functional
\begin{equation}
    W[\eta] = -i \ln \int\! \mathcal{D}\Phi \, \exp\bigg(iS[\Phi] + \frac{i}{4} \int_\mathcal{C}\! dt \, \sum_{n\alpha} \eta^\alpha_n \varphi^T_n K^\alpha \varphi_n \bigg) \\
\end{equation}
via functional dereivative with respect to the source field $\eta^\alpha_n$ in the limit where $\eta^\alpha_n$ vanishes, i.e.,
\begin{equation}
    \braket{s^\alpha_n(t) s^\beta_{n'}(t')} = -i \frac{\delta^2 W[\eta]}{\delta \eta^\alpha_n(t) \delta \eta^\beta_{n'}(t')} \bigg|_{\eta=0}
\end{equation}
The source term can be absorbed into the effective Hamiltonian, so that $\tilde H_n[\eta] = \frac{1}{4}\sum_\alpha (h^\alpha +\lambda^\alpha_n + \Lambda^\alpha_n + \eta^\alpha_n)  K^\alpha$. Since the functional integral is over all configurations of the fields, it is invariant under the constant shift $\Lambda \rightarrow \Lambda - \eta$. Such shift produces an action identical to the the one in Eq.~\eqref{eq:finalAction}, except that the last term on the RHS becomes
\begin{equation}
    \frac{1}{4} \int_\mathcal{C} \! dt \, \sum (\Lambda^\alpha_n - \eta ^\alpha_n)[J^{-1}]^{\alpha\beta}_{nn'} (\Lambda^\beta_{n'}- \eta ^\alpha_n).
\end{equation}
Now taking the functional derivative with respect to source $\eta$ generates EVs of the mean field $\Lambda$. It turns out that
\begin{equation}\label{eq:twospin}
    \braket{s^\alpha_n(t) s^\beta_{n'}(t')} = \frac{i}{4} \check M^{\alpha\beta}_{nn'} (t,t'),
\end{equation}
where we used Eq.~\eqref{eq:mflocal} to isolate the time-local part of the mean field propagator. Instead of shifting the mean field $\Lambda^\alpha_n$, the bath field $\lambda^\alpha_n$ can also be shifted, potentially leading to an expression for the two-spin correlator in terms of the bath propagator $D^\alpha_n$. However, such approach is made difficult by requiring to carefully invert convolutions with the bath kernel $\Xi^\alpha_n$ on the SK contour. Thus, we compute two-spin correlators exclusively via Eq.~\eqref{eq:twospin}.

In single-spin systems, such as the spin-boson model [Eq.~\eqref{eq:sbhamilton}], there is no mean field propagator. Still, the two-spin correlator $\braket{s^\alpha (t) s^\beta(t')}$ at the same site but for different times and spin components can be obtained by introducing a replica spin and bath with identical initial states. By weakly coupling the spin and its replica ferromagnetically, the two-spin correlator between the spin and its replica mimics the original same-site two-spin correlator.

\subsection{Numerical implementation of SKFT+2PI equations}\label{sec:numerical}

The equations of motion produced by SKFT+2PI form an integro-differential system of the Volterra type~\cite{Meirinhos2022, Blommel2024} that must be integrated carefully due to the self-consistent interdependence between 14 functions. For this purpose, we discretize both time arguments $t$ and $t'$, so that all functions of two times can be considered matrices in these two arguments. The Keldysh and spectral components are symmetric and antisymmetric, respectively, under transposition and exchange of the two time arguments, i.e.,
\begin{equation}\label{eq:transpose}
O^{K/s} (t,t')=\pm (O^{K/s})^T(t',t). 
\end{equation}
Therefore, it suffices to compute and store all quantities at times $t'\leq t$.

For the numerical integration of the integro-differential system in Eq.~\eqref{eq:eomReal}, we implement a predictor-corrector algorithm which consists of two stages. In the first (predictor) stage, the RHSs of Eqs.~\eqref{eq:eomRealF} and~\eqref{eq:eomRealRho} are computed using all known quantities for all $t'$ and the current $t$. The results are used to predict the GFs of the Schwinger bosons at the next time $t+h$, i.e., $\tilde g^{K/s}_n(t+h,t') = g^{K/s}_n(t,t') + h\partial_t g^F_n(t,t')$ where $h$ is the time step, for all discrete $t'$. Predicting the diagonal time step $\tilde g^K_n(t+h,t+h)$ requires the derivative with respect to the second time argument, $\partial_{t'}g^K_n(t,t')$, obtained from the transpose of Eq.~\eqref{eq:eomRealF} and the symmetry properties of the nonequilibrium GFs in Eq.~\eqref{eq:transpose}. The time-diagonal of $g^s_n(t,t)=-i\langle[\varphi^b_n(t), \varphi^a_n(t)]\rangle = i K^0$ is fixed by the equal-time commutation relations of the Schwinger bosons. With the predicted values for the GFs of the Schwinger bosons $\tilde g^{K/s}_n(t+h,t')$, the value of all the other functions at the next time step can be predicted in the order shown in Fig.~\ref{fig:flowchart}. All integrals appearing on the RHSs of Eqs.~\eqref{eq:eomReal} are computed with the trapezoid method.

In the second (corrector) stage of the predictor-corrector algorithm, the predicted quantities are used to recompute the RHSs of Eqs.~\eqref{eq:eomRealF} and~\eqref{eq:eomRealRho}. The new value for the GFs of the Schwinger bosons is then obtained from averaging the RHSs of Eqs.~\eqref{eq:eomRealF} and~\eqref{eq:eomRealRho} obtained in the prediction and correction stages, i.e.,
\begin{align}
g^{K/s}_n(t+h,t') &= g^{K/s}_n(t,t') \\
&+ \frac{h}{2}\big[\partial_t g^{K/s}_n(t,t') + \partial_t \tilde g^{K/s}_n(t+h,t')\big]. \nonumber
\end{align} 
Using the corrected $g^{K/s}_n(t+h,t')$, all other quantities are recalculated following the same order as in the predictor stage [Fig.~\eqref{fig:flowchart}]. Additional details of this type of predictor-corrector scheme can be found in Ref.~\cite{Schuckert2018}.

\begin{figure}
    \centering
    \includegraphics[width=\columnwidth]{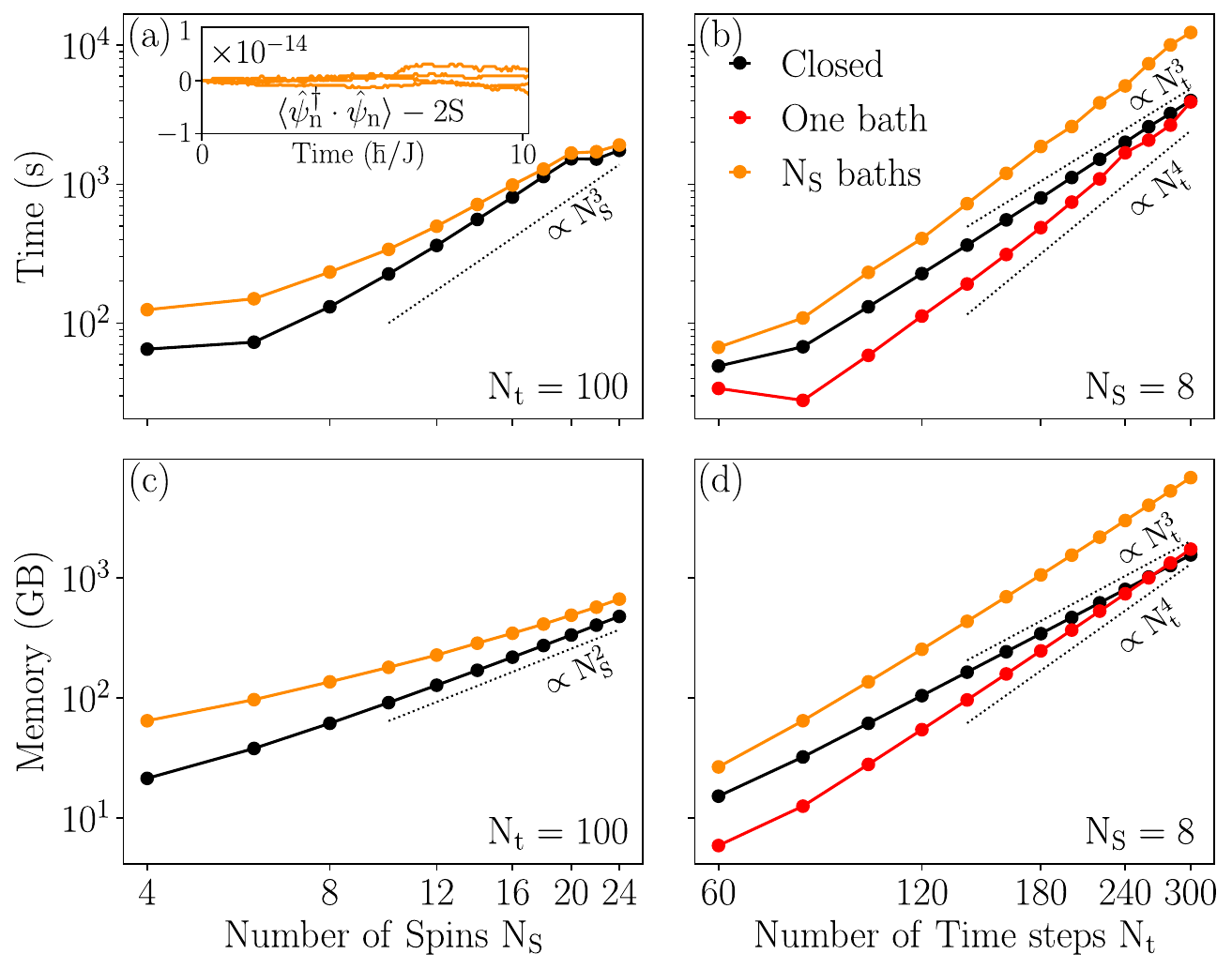}
    \caption{(a)--(b) Elapsed time and (c)--(d) memory consumed by our numerical implementation of the predictor-corrector algorithm for solving the SKFT+2PI Eqs.~\eqref{eq:eomReal}(a)--(n) as a function of the number of spins $N_S$ (left column of panels) or the number of time steps $N_t$ (right column of panels). The black lines are for a closed AF spin chain in a transverse magnetic field [Eq.~\eqref{eq:chainhamiltonS}]; the red ones are for the spin-boson model [Eq.~\eqref{eq:sbhamilton}]; and the orange ones are for the spin-chain-boson model with $N_S$ baths [Eq.~\eqref{eq:chainhamilton}] where \textit{each} spin couples to three independent baths. The inset in panel (a) shows numerical deviations from the EV of the Schwinger boson constraint [Eq.~\eqref{eq:constraintEV}] in the case of a spin-chain-boson model with $N_S=4$. }
    \label{fig:scaling}
\end{figure}

In order to estimate the numerical cost of solving the equations of motion produced by SKFT+2PI, we evolved one closed system and two open systems using our implementation of the predictor-corrector algorithm. For the closed system, we choose an AF spin chain subjected to a transverse field along the $x$-axis [Eq.~\eqref{eq:chainhamiltonS}]. For the open systems, we evolve the spin-boson model [Eq.~\eqref{eq:sbhamilton}] and the spin-chain-boson-model with an independent bosonic bath for each spin  [Eq.~\eqref{eq:chainhamilton}, but the first sum in Eq.~\eqref{eq:chainhamiltonB} is over $n=1,2,\ldots, N_S$]. As a function of the number of spins $N_S$, the elapsed time for both closed and open systems scales as $\propto N_S^3$. This is due to the largest matrix being the propagator $\check M^{\alpha\beta}_{nn'}(t,t')$ (Table~\ref{tab:symbols}), which can be reshaped into a $3N_S \times 3N_S$ matrix. Concomitantly, the total memory allocated scales as $\propto N_S^2$. On the other hand, as a function of the number of time steps $N_t$, both the elapsed time and the memory scale as $\propto N_t^3$ for the closed system, and as $\propto N_t^4$ for the open systems. This is because evolving functions of two times $t,t'$ for $N_t$ time steps requires computing and storing such functions at $\propto N_t^2$ pairs of times. Moreover, computing each one of those values requires performing single (for the closed system) or double (for the open systems) time integrals, which scale as $\propto N_t$ and $\propto N_t^2$, respectively, therefore producing the observed scaling.

\subsection{Initial conditions and the Schwinger boson constraint}\label{sec:initial}

Initial conditions must be given to begin the predictor-corrector algorithm. Within the SKFT+2PI approach, initial states must be described by a Gaussian density operator~\cite{Berges2015, Schuckert2018, Babadi2015}, i.e., only the EVs and the connected nonequilibrium GFs of the fields are nonzero. In this work, we consider separable initial states described by a density operator   that is a tensor product
\begin{equation}
    \hat{\rho}(t=0) = \bigotimes_\nu \hat \rho^B_\nu \bigotimes_n \hat{\rho}^S_n.
\end{equation}
Here, $\hat \rho^B_\nu$ is the thermal density operator  of bosonic bath $\nu$ at temperature $T_\nu$, and $\hat\rho^S_n$ is a Gaussian density operator for spin $n$. In our numerical implementation of the SKFT+2PI equations, these initial conditions are achieved by setting the bath propagator as $D^{K/s}(0,0)=2\Xi^{K/s}(0,0)$, as well as the GFs of the Schwinger bosons as $i g^K_n(0,0) = \sum_\alpha K^\alpha\braket{\hat{\sigma}^\alpha_n(0)} + S +\frac{1}{2}$ and $g^s_n=iK^0$.

Due to the symmetries of  the SK action [Eq.~\eqref{eq:finalAction}], the total number of bosons per site is conserved~\cite{Schuckert2018}. Therefore, if the initial conditions satisfy the Schwinger boson constraint in Eq.~\eqref{eq:constraint}, it will also be satisfied at \textit{all} later times. This is in contrast to the usage of Schwinger bosons in imaginary time calculations, which require enforcing the constraint via Lagrange multipliers~\cite{Zhang2022, Auerbach1994}. The Gaussian initial conditions we employ enforce that the EV of the constraint
\begin{equation}
    \left\langle\hat a_n^{(1)\dagger} \hat a_n^{(1)} + \hat a^{(2)\dagger}_n \hat a^{(2)}_n\right\rangle = 2S,
\end{equation}
holds at all times. Deviations from this equality are typically less than $10^{-14}$ in our numerical implementation [inset of Fig.~\ref{fig:scaling}(a)]. However, the original Schwinger boson constraint [Eq.~\eqref{eq:constraint}] relates \textit{operators}, which implies infinite additional constraints in terms of EVs, namely,
\begin{subequations}\label{eq:constraintEV}
\begin{align}
    \left\langle \left(\hat a_n^{(1)\dagger} \hat a_n^{(1)} + \hat a^{(2)\dagger}_n \hat a^{(2)}_n \right)^2 \right\rangle &= (2S)^2, \label{eq:constraintEVb} \\
    \left\langle \left(\hat a_n^{(1)\dagger} \hat a_n^{(1)} + \hat a^{(2)\dagger}_n \hat a^{(2)}_n \right)^3\right\rangle &= (2S)^3 \label{eq:constraintEVc},\\
    &\vdots \nonumber
\end{align}
\end{subequations}
Because these higher-order EVs are zero for an initial Gaussian density operator, the constraints in Eqs.~\eqref{eq:constraintEV} are not satisfied at any time $t\geq 0$. Thus, artifactual virtual processes outside the physical finite-size Hilbert space of spins can contribute (see Sec.~\ref{sec:sbresults}) to the discrepancy between SKFT+2PI and numerically (quasi)exact benchmark results. It is worth noting that problems posed by the Gaussian initial state are unrelated to the truncation of the 2PI effective action. In fact, \textit{if} one were able to provide the full correct initial state, all Schwinger boson constraints derived from the operator identity in Eq.~\eqref{eq:constraint} would be satisfied despite truncating the diagrammatic expansion. However, usage of arbitrary initial states for time evolution via SKFT is an unsolved problem, despite many proposed remedies~\cite{Garny2009, vanLeeuwen2013, Carrington2016, Chakraborty2019, Kaeding2023}.

\section{Results and discussion}\label{sec:results}

\subsection{Dynamics of semiclassical spins}\label{sec:semiclassical}

To understand the physical meaning of SKFT+2PI Eqs.~\eqref{eq:eomReal}, we warm up in this section by discussing semiclassical spin dynamics~\cite{ReyesOsorio2024, Verstraten2023, Anders2022}, which is easier to decipher than fully quantum dynamics of spin-boson (Sec.~\ref{sec:sbresults}) and spin-chain-boson (Sec.~\ref{sec:chainresults}) models. Such dynamics is obtained from SKFT+2PI by neglecting even the $1/N^0$ contributions to the 2PI effective action, i.e., $\Gamma_2 \equiv 0$ in Eq.~\eqref{eq:2piAction} is set to zero. Under this approximation, all SEs vanish, and the Eqs.~\eqref{eq:eomReal} of motion decouple. Using Eq.~\eqref{eq:spinEV} for the EVs of the spins and Eq.~\eqref{eq:eomRealF} for the evolution of the Keldysh GF of the Schwinger bosons $g^K$ yields
\begin{equation}
    \partial_t\braket{s^\alpha_n}(t) = \sum_{\beta\gamma} \epsilon_{\alpha\beta\gamma} (h^\beta_n + \lambda^\beta_n + \Lambda^\beta_n)\braket{s^\alpha_n}(t),
\end{equation}
where $\epsilon_{\alpha\beta\gamma}$ is the Levi-Civita symbol. By defining the vector of spin EVs, $\mathbf{S}_n\equiv (\braket{s^x_n}, \braket{s^y_n}, \braket{s^z_n} )$, and by using Eqs.~\eqref{eq:fields} for the fields $\lambda^\beta_n$ and $\Lambda^\beta_n$, the time evolution $\mathbf{S}_n(t)$ is given by
\begin{equation}\label{eq:newllg}
    \partial_t \mathbf{S}_n(t) = \mathbf{h}^{\rm eff}_n \times \mathbf{S}_n(t) - \mathbf{S}_n (t) \times \int_0^t \! dt' \, \eta_n(t,t') \cdot \mathbf{S}_n(t'),
\end{equation}
where the effective magnetic field is \mbox{${h}^{\alpha \, \rm eff}_n = h_n^\alpha + \Lambda^\alpha_n$} and the time-retarded non-Markovian memory kernel. Note that Eq.~\eqref{eq:newllg} is of the Landau-Lifshitz type~\cite{Landau1935}, but it is ``extended'' by having non-Markovian kernel \mbox{$\eta_n^{\alpha\beta}(t,t') = \delta^{\alpha\beta}\Xi^{\alpha s}_n(t,t')$} due to effects of the bosonic baths. Similar equations have been recently derived~\cite{Anders2022, Verstraten2023, Bajpai2019, ReyesOsorio2024, ReyesOsorio2025, Bhattacharjee2012}, including via the minimization of the SK action with respect to quantum fluctuations~\cite{ReyesOsorio2024, Verstraten2023}. Thus, our SKFT+2PI formalism further  {\em justifies} such derivations by explaining that they  utilize {\em only} the first three terms on the RHS of Eq.~\eqref{eq:2piAction}  containing tree-level or classical [1st term on the RHS of Eq.~\eqref{eq:2piAction}] and  one-loop or semiclassical [2nd and 3rd terms on the RHS of Eq.~\eqref{eq:2piAction}] diagrams of the 2PI effective  action~\cite{Fauth2021}. If multiple spins are allowed to couple to a single bath, the non-Markovian kernel in Eq.~\eqref{eq:newllg}, besides being nonlocal in time, also becomes nonlocal in space~\cite{ReyesOsorio2024, Bhattacharjee2012}.

\begin{figure}
    \centering
    \includegraphics[width=\columnwidth]{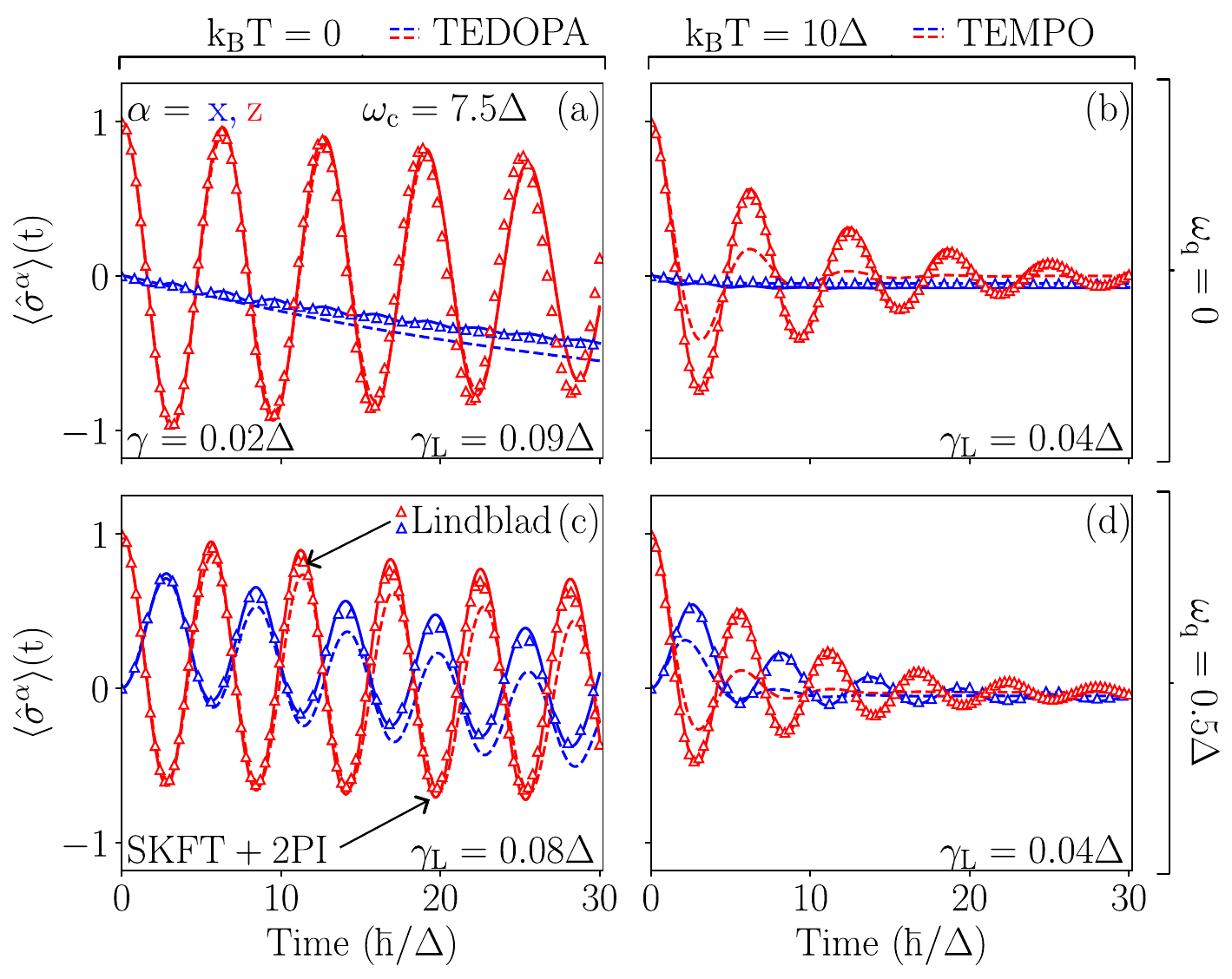}
    \caption{Time dependence of spin expectation values $\langle \hat{\sigma}^\alpha \rangle$ of the spin-boson model [Eq.~\eqref{eq:sbhamilton}] computed from our SKFT+2PI (solid lines) approach vs. standard Lindblad QME (triangles) or TN methods (dashed lines) with bath cutoff frequency $\omega_c=7.5\Delta$ in the \textit{weak} system-bath coupling regime, $\gamma\ll\Delta$, where the system dynamics is expected~\cite{Nathan2020} to be {\em Markovian}~\cite{Clos2012, Wenderoth2021} for the chosen Ohmic bath [$s=1$ in Eq.~\eqref{eq:spectral}]. Panels in different columns and rows use different values of temperature $T$ and two-level splitting $\omega_q$, respectively. For $k_B T=0$, the chosen TN method is TEDOPA, while for $k_BT=10\Delta$, we use TEMPO (see Secs.~\ref{sec:tensor} and~\ref{sec:results} for details). The SKFT+2PI and TN-computed benchmark results follow each other closely for the chosen $\gamma=0.02\Delta$, while the system-bath coupling in Eqs.~\eqref{eq:lindblad} and~\eqref{eq:jumpOps} for the Lindblad QME, $\gamma_{\rm L}$, must be adjusted by increasing it to match the other two calculations (this points to artifacts of the Lindblad QME derived~\cite{Breuer2007} for the spin-boson model).}
    \label{fig:lindblad}
\end{figure}

\begin{figure}
    \centering
    \includegraphics[width=\columnwidth]{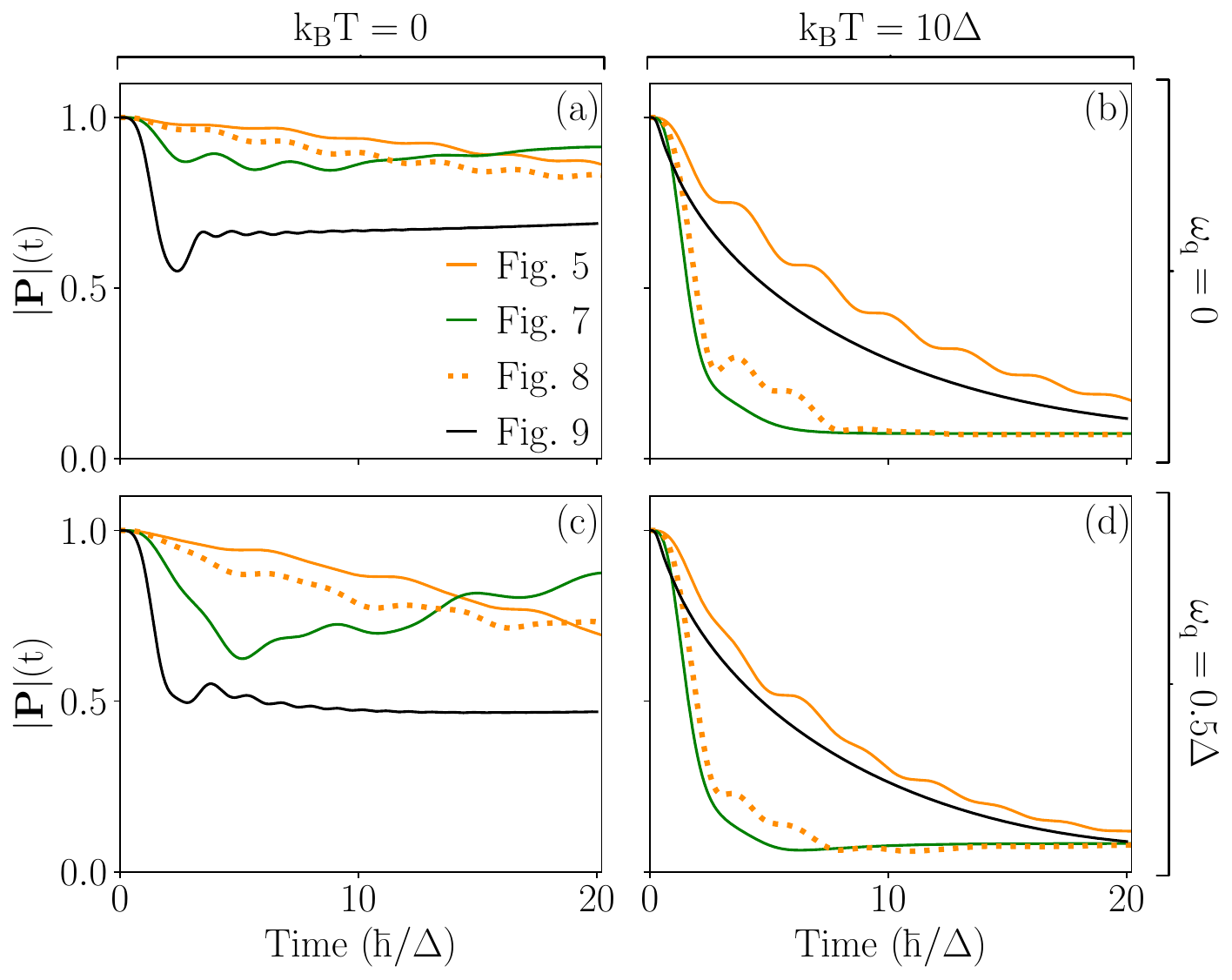}
    \caption{Time dependence of purity $|\mathbf{P}|=\sqrt{\sum_\alpha \langle \hat \sigma_\alpha \rangle^2}$ of mixed quantum state of spin $S=1/2$ from SKFT+2PI-computed curves in Figs.~\ref{fig:lindblad}--\ref{fig:strong}. Orange, green, and black solid lines are for the case of an Ohmic bath [Fig.~\ref{fig:lindblad}, Fig.~\ref{fig:heom} and Fig.~\ref{fig:strong}], while orange dotted line is for the case of a sub-Ohmic bath [Fig.~\ref{fig:subohmic}].}
    \label{fig:coherence}
\end{figure}

The spectral component of the bath kernel $\Xi^{\alpha s}_n(t,t')$, which yields the non-Markovian kernel within Eq.~\eqref{eq:newllg}, is computed for the spectral density in Eq.~\eqref{eq:spectral} for arbitrary parameter $s$ via Eqs.~\eqref{eq:bathkernel} and~\eqref{eq:freeBathGF} to give
\begin{equation}
    \Xi^{\alpha s}_n(t,t') = -\frac{\gamma^\alpha_n \omega_c^2}{\pi} \frac{\Gamma(1+s) \sin[(1+s) \tan^{-1}(\omega_c\tau)]}{[1+(\omega_c\tau)^2)]^{(1+s)/2}}.
\end{equation}
Here, $\tau\equiv t-t'$, and $\Gamma(x)$ is the Gamma function. The slow algebraic decay of $\Xi^{\alpha s}_n(\tau)$ is linked to non-analyticities of the spectral density~\cite{Chakraborty2018}, namely, the second derivative of $\mathcal{J}(\omega)$ in Eq.~\eqref{eq:spectral} is not defined at $\omega=0$. Such power law decay of the bath kernel means that the contribution of states in the far past to the dynamics of open quantum systems can persist beyond other relevant time scales. For instance, even if the system-bath coupling $\gamma$ is small, a low cutoff frequency $\omega_c$ ensures that the power law tails of the bath are relevant, and the dynamics \textit{remains non-Markovian}. 

Nevertheless, Markovian dynamics of the extended LLG Eq.~\eqref{eq:newllg} can be recovered in the limit $\omega_c\rightarrow \infty$ of an Ohmic bath $s=1$~\cite{Verstraten2023, Anders2006} for which the bath kernel becomes local in time $\Xi^{\alpha s}_n (t,t') = \gamma_n^\alpha \partial_t \delta(t-t')$. In this limit, the non-Markovian Eq.~\eqref{eq:newllg} simplifies into
\begin{equation}\label{eq:llg}
     \partial_t \mathbf{S}_n(t) = \mathbf{h}^{\rm eff}_n \times \mathbf{S}_n(t) - \gamma_n \mathbf{S}_n (t) \times \partial_t \mathbf{S}_n(t),
\end{equation}
where we also assume isotropic spin-bath couplings, for simplicity. Equation~\eqref{eq:llg} is the standard Landau-Lifshitz-Gilbert equation~\cite{Landau1935}, in which damping term (second on the RHS) is in the Gilbert form~\cite{Saslow2009}. Considering the  parameter $s\neq 1$ transmutes the time derivative in the damping term of Eq.~\eqref{eq:llg} into a fractional derivative~\cite{Verstraten2023}. 

\subsection{Dynamics of the spin-boson model}\label{sec:sbresults}

\begin{figure}
    \centering
    \includegraphics[width=\columnwidth]{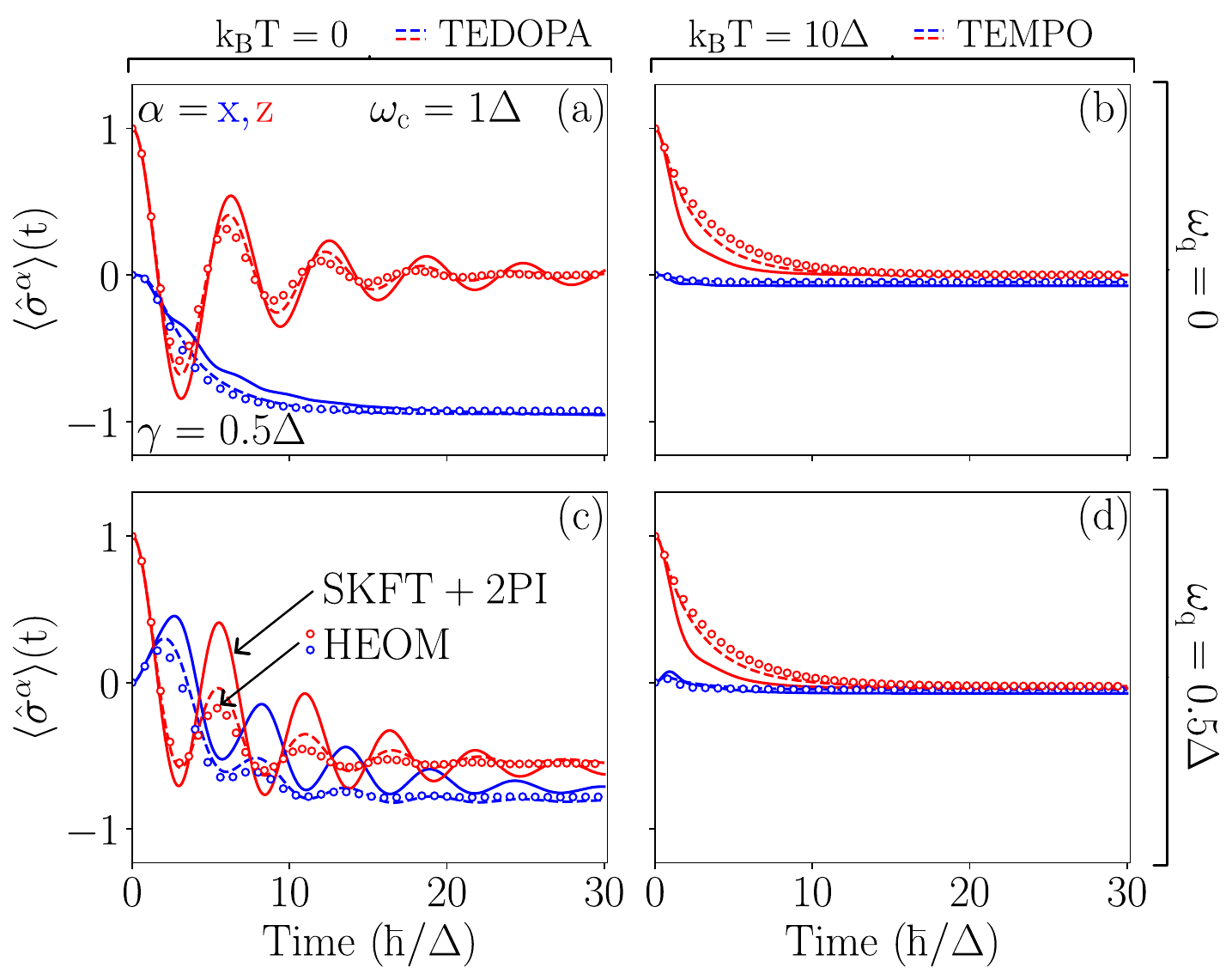}
    \caption{The same information as in Fig.~\ref{fig:lindblad}, but for  bath cutoff frequency $\omega_c=1\Delta$ and in the \textit{strong} system-bath coupling regime, \mbox{$\gamma=0.5\Delta$}, where the system dynamics is expected to be {\em non-Markovian}~\cite{Clos2012, Wenderoth2021} for the chosen~\cite{Nathan2020}  Ohmic bath [$s=1$ in Eq.~\eqref{eq:spectral}]. Different columns and rows of panels use different values of temperature $T$ and two-level splitting $\omega_q$ [Eq.~\eqref{eq:sbhamilton}], respectively. Note that standard HEOM calculations~\cite{Tanimura2020} cannot be conducted at $k_B T=0$ temperature, so in the left column of panels we use higher temperature $k_B T=0.1\Delta$ in HEOM calculations instead of $k_B T=0$ [as marked on the top of panel (a)] employed in SKFT+2PI and TN-based calculations.}
    \label{fig:heom}
\end{figure}

In this Section, we present dynamics of spin EV in the spin-boson model [Eq.~\eqref{eq:sbhamilton}] using the Ohmic [$s=1$ in Eq.~\eqref{eq:spectral}] and sub-Ohmic baths [$s=0.5$ in Eq.~\eqref{eq:spectral}]. We start with the analysis of the Markovian regime in Fig.~\ref{fig:lindblad} for which we employ small $\gamma=0.02\Delta$ and high $\omega_c=7.5\Delta$ to ensure entering such a regime~\cite{Clos2012}. The Markovian nature of the results in Fig.~\ref{fig:lindblad} is signified by the irreversible decay of purity [orange solid line in Fig.~\ref{fig:coherence}] of the mixed quantum state of spin $S=1/2$. Since the spin density operator $\hat\rho^S$ for $S=1/2$ and the Bloch vector $\mathbf{P}=(P^x, P^y, P^z)$ are in one-to-one correspondence~\cite{Ballentine2014}
\begin{equation}\label{eq:densityMatrix}
    \hat\rho = \frac{1}{2}\left(\hat I + \sum_\alpha P^\alpha \hat\sigma^\alpha \right),
\end{equation}
where $\hat I$ is the unit operator in the spin space, we use $|\mathbf{P}|$ as a measure of the state purity. The standard purity ${\rm Tr}\hat\rho^2$ is a function of $|\mathbf{P}|$, where $|\mathbf{P}|=1$ signifies a fully coherent or pure quantum state of spin $S=1/2$, while $0\leq |\mathbf{P}|<1$ denotes mixed quantum states. In Fig.~\ref{fig:lindblad}, we find excellent agreement between SKFT+2PI (solid lines) and Lindblad-QME-computed results (triangles). However, such a match is \textit{ensured only} by adjusting the system-bath coupling in the Lindblad QME, thereby pointing to an artifact of the standard equation for the spin-boson model~\cite{Breuer2007}. This is because our SKFT+2PI results independently and closely match the results of TN calculations (dashed lines in Figs.~\ref{fig:lindblad} and~\ref{fig:heom}) employing the same $\gamma$. Our TN calculations for the $k_B T=0$ case used TEDOPA~\cite{deVega2015, Chin2010, Ye2021}, whereas for the higher temperature $k_BT=10\Delta$ we switched to TEMPO~\cite{Strathearn2018, Fux2024}. Beyond the Markovian regime, the Lindblad QME cannot capture the memory effects of the bath, which can cause the revival of quantum properties. Such revival is exemplified by the purity $|\mathbf{P}|$ of the mixed quantum state of spin initially decaying in Figs.~\ref{fig:coherence}(a) and~\ref{fig:coherence}(c), as the signature of decoherence~\cite{Joos2003}, but later increasing towards $|\mathbf{P}|=1$ of the pure state at $t=0$ as the signature of recoherence~\cite{Breuer2016}. 

\begin{figure}
    \centering
    \includegraphics[width=\columnwidth]{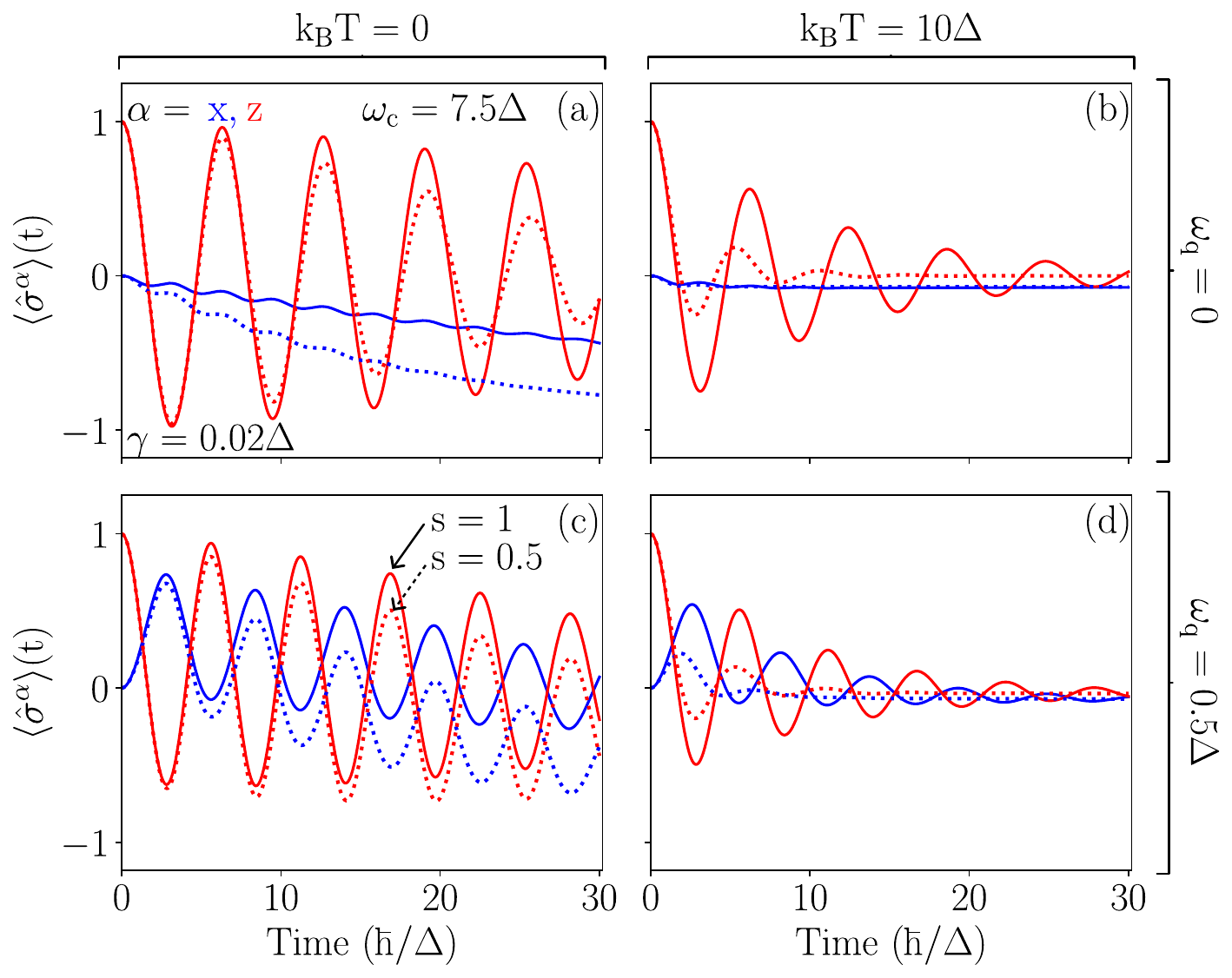}
    \caption{Time dependence of spin expectation values $\langle \hat{\sigma}^\alpha \rangle$ computed from SKFT+2PI for bosonic bath with Ohmic (solid lines) or sub-Ohmic (dotted lines) spectral density, i.e., $s=1$ or $s=0.5$ in Eq.~\eqref{eq:spectral}, respectively. Other parameters are the same as in the weak system-bath coupling regime of Fig.~\ref{fig:lindblad}. Note that solid lines are identical to solid lines in Fig.~\ref{fig:lindblad}, which are plotted here for easy comparison.}
    \label{fig:subohmic}
\end{figure}

In the non-Markovian regime of Fig.~\ref{fig:heom}, we replace benchmarking via the Lindblad QME with HEOM and TN-based benchmarks. The non-Markovian regime is induced by using a strong system-bath coupling \mbox{$\gamma=0.5\Delta$} and low cutoff frequency \mbox{$\omega_c=\Delta$}, while keeping the Ohmic bath as in Fig.~\ref{fig:lindblad}. We find that, for zero temperature, the TEDOPA results [dashed lines in Figs.~\ref{fig:heom}(a) and~\ref{fig:heom}(c)] follow closely those from HEOM [circles in Fig.~\ref{fig:heom}], on the proviso that we use slightly higher temperature $k_B T=0.1\Delta$ in the HEOM calculations. The necessity for such an \textit{ad hoc} fixing of standard HEOM calculations stems from their \textit{inability} to handle $k_B T=0$ limit~\cite{Xu2022,Xu2023,Xu2023a}. The spin EV computed by SKFT+2PI is capable of tracking both benchmark results, but it appears as if its  damping is slightly smaller [compare solid lines from SKFT+2PI to circles from HEOM and dashed lines from TEDOPA calculations in Fig.~\ref{fig:heom}(c)]. 

\begin{figure}
    \centering
    \includegraphics[width=\columnwidth]{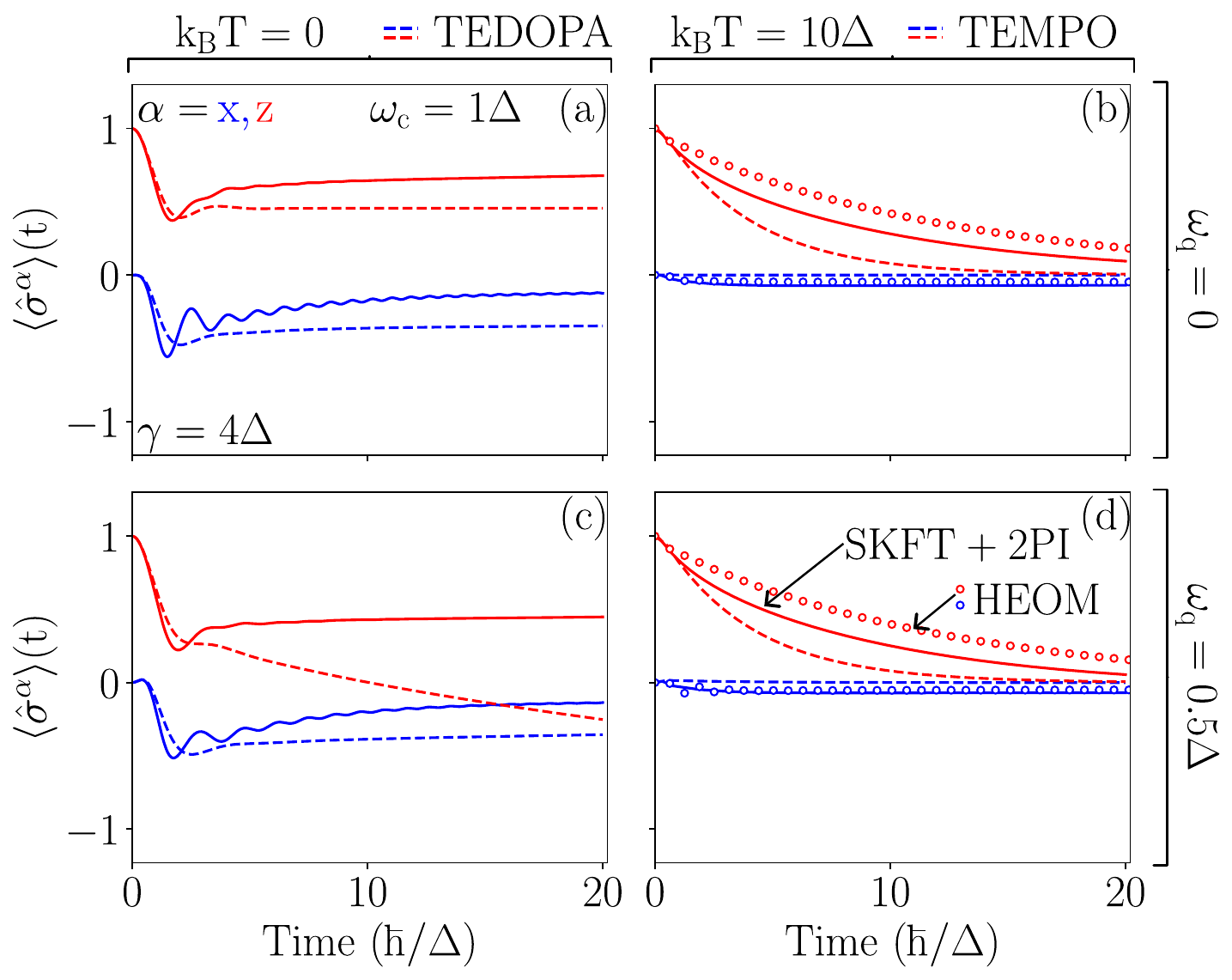}
    \caption{The same information as in Fig.~\ref{fig:heom}, but for  bath cutoff frequency $\omega_c=1\Delta$ and in the \textit{ultrastrong} system-bath coupling regime, \mbox{$\gamma=4\Delta$}, where the system dynamics is expected to be {\em highly non-Markovian}~\cite{Clos2012, Wenderoth2021} for the chosen~\cite{Nathan2020} Ohmic bath [$s=1$ in Eq.~\eqref{eq:spectral}]. The ultrastrong system-bath coupling induces localization of the spin for $k_B T=0$, i.e., $\braket{\hat\sigma^z}(t)$ plateaus at a finite noninteger value. This non-perturbative behavior of a plateau reached in the long-time limit is captured by SKFT+2PI (solid lines), but the asymptotic value defining the plateau differs from the TN-computed benchmark (dashed lines) at low temperatures.}
    \label{fig:strong}
\end{figure}

Furthermore, Fig.~\ref{fig:subohmic} demonstrates the ability of our SKFT+2PI formalism to treat various other parameter regimes of the system and bath, such as the case of zero temperature and sub-Ohmic bath that is considered particularly challenging~\cite{Wang2010,Xu2023,Bulla2003,Anders2006,Anders2007,Vojta2012}. For this purpose, we compute from SKFT+2PI the spin EV for a sub-Ohmic bath [$s=0.5$ in Eq.~\eqref{eq:spectral}] while using the same values of other parameters as in the weak system-bath coupling regime of Fig.~\ref{fig:lindblad} for the sake of comparing Ohmic vs. sub-Ohmic cases. The results in Fig.~\ref{fig:subohmic} show faster decrease of the spin EV when the bath is sub-Ohmic. However, the purity in the sub-Ohmic case [orange dotted line in Fig.~\ref{fig:coherence}] \textit{does not} decay monotonically as in the case of the Markovian regime for the Ohmic bath [orange solid line in Fig.~\ref{fig:coherence}]. Instead, at zero temperature it saturates at a finite value, akin to the Ohmic non-Markovian case [green and black lines in Fig.~\ref{fig:coherence} obtained using SKFT+2PI from Fig.~\ref{fig:heom} and Fig.~\ref{fig:strong}]. Moreover, at high temperature, the purity of spin state in the sub-Ohmic regime closely resembles the time evolution of it in the strong system-bath coupling (i.e., non-Markovian) regime, except for small revivals [orange dotted curve in Figs.~\ref{fig:coherence}(b),(d)] at intermediate time scales. Thus, Fig.~\ref{fig:coherence} illustrates the difficulties~\cite{Wang2010,Xu2023,Bulla2003,Anders2006,Anders2007,Vojta2012} posed by the sub-Ohmic case because of the skew towards low bath frequencies [Fig.~\ref{fig:fig0}(a)] which enhance the memory effects of the bath. These features make it possible for the non-Markovian regime to emerge \textit{despite} weak system-bath coupling. 

\begin{figure}
    \centering
    \includegraphics[width=\linewidth]{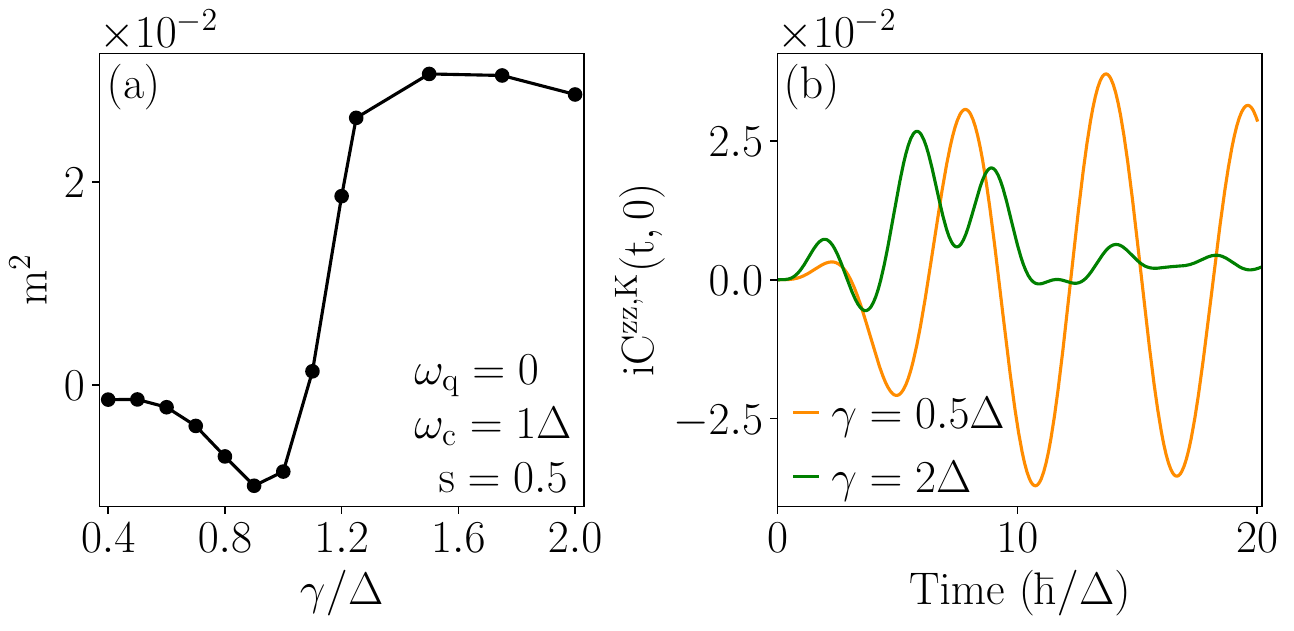}
    \caption{(a) Order parameter [Eq.~\eqref{eq:orderParam}] of the quantum phase transition~\cite{LeHur2008, DeFilippis2020} of the spin-boson model as a function of the system-bath coupling constant $\gamma$. The transition is signified by a kink at the critical coupling $\gamma_c \approx \Delta$~\cite{DeFilippis2020}. (b) Multitime two-spin correlator, at time $t$ and the initial time $t=0$, for $\gamma$ values in the delocalized (orange line) and localized phase (green lines), respectively.}
    \label{fig:qpt}
\end{figure}

Finally, we examine the non-Markovian regime brought by ultrastrong system-bath coupling $\gamma=4\Delta$ [Fig.~\ref{fig:strong}], while keeping the remaining parameters the same as in Fig.~\ref{fig:heom}. At zero temperature of this regime, the spin-boson transition to a localized phase occurs~\cite{LeHur2008, DeFilippis2020}. It is characterized by long-range correlations in time and $\braket{\hat\sigma^z}(t)$ saturating at a finite value despite the presence of the dissipative environment and regardless of two-level splitting $\omega_q$. Such saturation is observed in the stationary state of the spin EV obtained with both TEDOPA and our SKFT+2PI, demonstrating the ability of the latter to capture highly non-perturbative effects. The quantum phase transition can be further probed by computing the order parameter $m^2$~\cite{LeHur2008, DeFilippis2020}, given by the time integral 
\begin{equation}\label{eq:orderParam}
m^2 = \int \! dt \, iC^{zz, K}(t,0),
\end{equation}
of the Keldysh component of the connected two-spin correlator \mbox{$C^{\alpha\beta}_{nn'}(t,t') = i\braket{\sigma^{\alpha}_n(t) \sigma^{\beta}_{n'}(t')}$}, as obtained via Eq.~\eqref{eq:keldyshgf}. Such an order parameter is zero in the delocalized phase [Figs.~\ref{fig:lindblad},~\ref{fig:heom} and~\ref{fig:subohmic}], but it abruptly acquires a finite value in the localized phase [Fig.~\ref{fig:strong}]. The two-spin correlator extracted from SKFT+2PI on the delocalized side of the transition oscillates as a function of time [orange line in Fig.~\ref{fig:qpt}(b)], whereas it is almost always positive on the localized side [green line in Fig.~\ref{fig:qpt}(b)]. Integrating the two-spin correlator over time for different system-bath couplings, the non-thermal critical coupling~\cite{Sachdev2011} is found to be $\gamma_c \approx\Delta$ in Fig.~\ref{fig:qpt}. In this calculation, we use $\omega_q=0$, $s=0.5$ and $\omega_c=\Delta$. 

Nevertheless, in the localized phase, SKFT+2PI predicts an incorrect long-time limit of the spin EV, especially for $\omega_q=0.5\Delta$ [Fig.~\ref{fig:strong}(c)]. We believe that such discrepancy stems from the initial conditions being restricted to a Gaussian density operator, since, in the localized phase, properties of the initial state are not erased by the bath. Such memory effects make the stationary spin EV particularly sensitive to any non-Gaussian initial correlations. These are not captured by SKFT+2PI, but are well described by TEDOPA. On the other hand, SKFT+2PI performs much better for ultrastrong coupling at high temperatures [Figs.~\ref{fig:strong}(b) and~\ref{fig:strong}(d)] than TN-based calculations like TEMPO, where the latter struggles to converge in the same regime. In fact, in this regime, SKFT+2PI results match the HEOM benchmark better than TEMPO results. Let us re-emphasize that the standard version of HEOM cannot~\cite{Xu2022,Xu2023,Xu2023a} handle zero temperature $k_BT=0$, which is essential for the transition to the localized phase, and so HEOM results are omitted from Figs.~\ref{fig:strong}(a) and~\ref{fig:strong}(c).

\subsection{Dynamics of the spin-chain-boson model}\label{sec:chainresults}

The spin-chain-boson model [Fig.~\ref{fig:fig0}(b)] employed in this Section is described by the Hamiltonian in Eq.~\eqref{eq:chainhamilton} using $N_S=4$ quantum spins $S=1/2$ with AF exchange coupling between the nearest neighbors \mbox{$J^{\alpha\beta}_{n,n+1} = J^{\alpha\beta}_{n,n-1} = J\delta^{\alpha\beta}$}. A set of three independent bosonic baths, one for each spin component, is coupled to each end of the chain, as needed for spins interacting with phonons in three-dimensional magnetic materials~\cite{Nemati2022,Hogg2024}. The temperature of the three baths coupled to spin $n=1$ is $k_BT_1=5J$, while the temperature of the three baths coupled to spin $n=4$ is $k_B T_4=0$. All other parameters are the same for simplicity, and the external magnetic field is chosen as $\mathbf{h} = (J,0,0)$.

\begin{figure}
    \centering
    \includegraphics[width=\columnwidth]{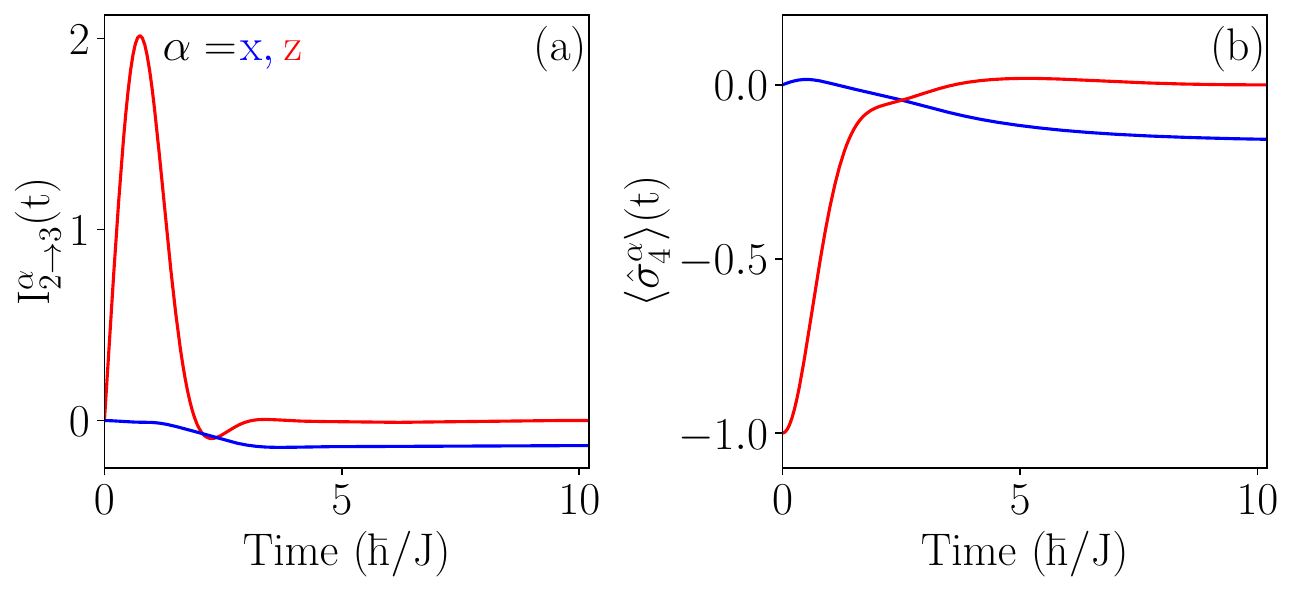}
    \caption{SKFT+2PI-computed time dependence of the Cartesian components of: (a) the bond spin current [Eq.~\eqref{eq:current}] between sites $n=2$ and $n'=3$; and (b) EV of spin at the right edge ($n=4$). The spin-chain-boson model considered in this calculation is AF, composed of $N_S=4$ spins and driven by temperature gradient, $k_B T_1=5J$ and $k_B T_4=0$, as illustrated in Fig.~\ref{fig:fig0}(b). Other parameters of the six baths employed are $\omega_c=3J$ and $\gamma^\alpha=0.5J$. 
    }
    \label{fig:transport}
\end{figure}

The dynamics of the spin $n=4$ EV is shown in Fig.~\ref{fig:transport}(b). Note that in this case we could not produce any benchmarks from either HEOM (because of too many spins) or from TN calculations (because of difficulties found in the packages~\cite{Fux2024} we employ when trying to handle multiple baths). The complexity of physics generated by multiple baths has been noticed previously~\cite{Bruognolo2014} even in the case of single spin. We start from the N\'eel ket $\ket{\uparrow\downarrow\uparrow\downarrow}$ in which $\braket{\hat\sigma^z_1}=\braket{\hat\sigma^z_3}=1$ and $\braket{\hat\sigma^z_2}=\braket{\hat\sigma^z_4}=-1$. The fact that dynamics of spin EV in Fig.~\ref{fig:transport}(b), as well as of EVs of other three spins not shown, tend to small values  $|\braket{\hat{\boldsymbol{\sigma}}_n}|\ll 1$ signifies nonzero entanglement of a mixed quantum. The entanglement can remain nonzero, despite the presence of a dissipative environment, due to the non-Markovian regime of open quantum system dynamics~\cite{GarciaGaitan2024}.

The open AF quantum spin chain with different temperatures of the baths connected at its boundaries~\cite{Landi2022} will also exhibit nonequilibrium spin and heat currents~\cite{Goehmann2022, MendozaArenas2013}. Since the total heat current requires three-spin correlators~\cite{Goehmann2022, MendozaArenas2013}, which are inaccessible from our current implementation of our SKFT+2PI formalism, we compute only spin current as an illustration. The $\alpha$-component of the bond spin current between two spins $n,n'$ is expressed as~\cite{Goehmann2022} 
\begin{equation}\label{eq:current}
    I^\alpha_{n\rightarrow n'}(t) = -2J\sum_{\beta\gamma}\epsilon_{\alpha\beta\gamma}\braket{\sigma^{\alpha}_n(t) \sigma^{\beta}_{n'}(t)}^K ,
\end{equation}
using the Keldysh component of two-spin correlators [Eq.~\eqref{eq:twospin} at equal times]. The time dependence of the bond spin currents $I^\alpha_{n\rightarrow n'}$, out of which we plot only $I^x_{2\rightarrow 3}$ and $I^z_{2\rightarrow 3}$ in Fig.~\ref{fig:transport}, is driven by the temperature difference between the baths at two different edges. The bond current $I^x_{2\rightarrow 3}$ has a finite value in the steady state because of the externally applied magnetic field $\mathbf{h}$ along the $x$-axis. On the other hand, $I^z_{2\rightarrow 3}$ is transiently nonzero but vanishes in the long-time limit. Transient oscillation of the currents within the spin chain can be long-lived due to finite-size effects. However, the decay of such oscillations in Fig.~\ref{fig:transport}(e) is accelerated by virtual processes outside of the physical Hilbert space. These processes can be caused by artifacts of resumming infinitely-many Feynman diagrams~\cite{PuigVonFriesen2010} or by neglected initial non-Gaussian correlations~\cite{Koksma2011}.

 

\subsection{Dynamics of closed quantum spins}\label{sec:closed}

\begin{figure}
    \centering
    \includegraphics[width=\columnwidth]{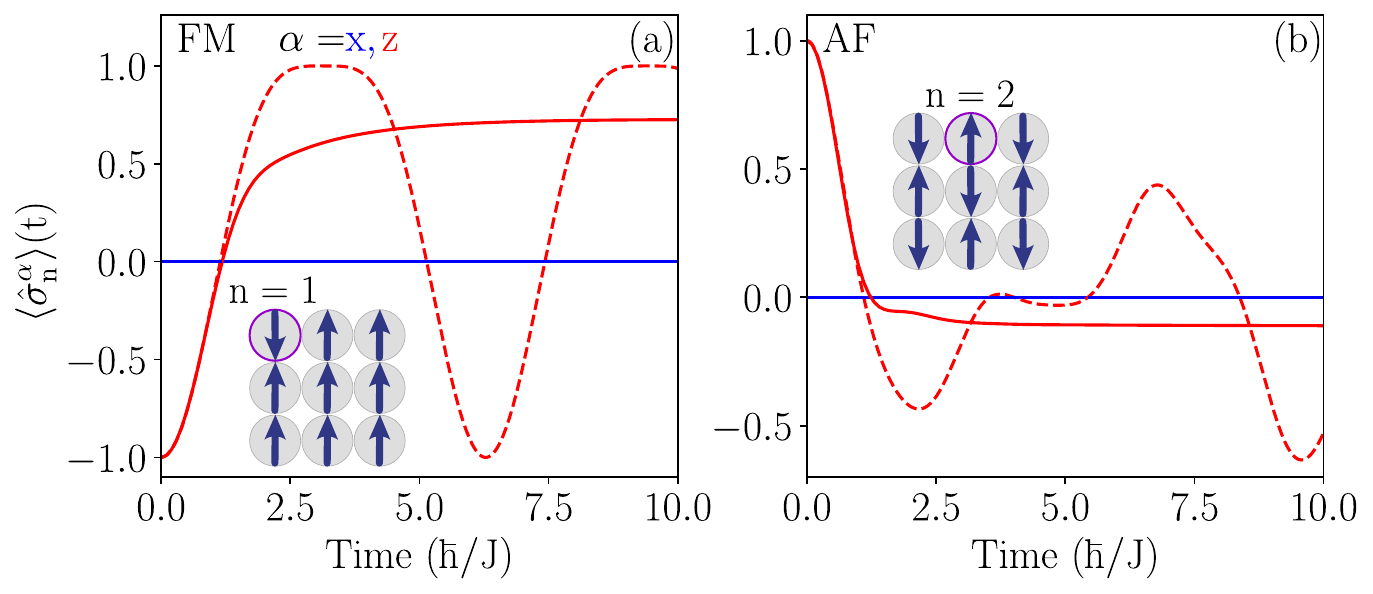}
    \caption{Time dependence of EV of a single spin $\braket{\hat \sigma^\alpha_n}(t)$, whose selection is marked in the insets, within a closed $3\times 3$ lattice of two-dimensional (a) FM or (b) AF cluster. The initial states are illustrated in the insets of each panel. The solid lines are computed by SKFT+2PI of Ref.~\cite{Schuckert2018} for closed quantum spin systems, while the dashed lines are obtained by exact diagonalization for time-dependent systems~\cite{Park1986}.}
    \label{fig:closed}
\end{figure}

Since our SKFT+2PI also describes the limiting case of a closed system of quantum spins, we also benchmark such calculations for the sake of completeness. Such closed quantum system dynamics is obtained by considering only the second diagram in Eq.~\eqref{eq:gamma2} due to spin-spin interaction. We recall that the time evolution of a closed system of quantum spins has already been studied by different flavors of SKFT+2PI formalism in Refs.~\cite{Schuckert2018} and~\cite{Babadi2015}, whose purpose is to enable handling of hundreds of spins in three spatial dimensions. In contrast, TN methods, which can handle large numbers of spins in low dimensions, fail in higher dimensions due to increased entanglement~\cite{Stoudenmire2012,Cirac2021,Patra2025}. However, examples studied in Refs.~\cite{Schuckert2018} and~\cite{Babadi2015} were not benchmarked against numerically exact dynamics, which can always be computed for sufficiently small systems. Therefore, in Fig.~\ref{fig:closed}, we plot the time evolution of EV of a selected spin within a $3\times 3$ cluster of FM [Fig.~\ref{fig:closed}(a)] or AF [Fig.~\ref{fig:closed}(b)] coupled quantum spins, as computed via SKFT+2PI (solid lines) or exact diagonalization for time-dependent systems~\cite{Park1986} (dashed lines). The comparison of trajectories in Fig.~\ref{fig:closed} shows that SKFT+2PI for closed quantum spin systems fails after a relatively short evolution time, appearing as if it is damped and, therefore, converging to a stationary state despite the \textit{absence} of a dissipative environment. This failure of previously developed SKFT+2PI for closed quantum systems~\cite{Schuckert2018, Babadi2015} is due to virtual processes outside the physical Hilbert space, as has also been pointed out in other problems involving fermionic~\cite{PuigVonFriesen2010} and bosonic~\cite{Rey2004} degrees of freedom. For fermionic systems, these virtual processes are thought to be found within the resummation of infinitely many Feynman diagrams, since diagrams of high enough order describe processes involving more fermions than are allowed by a finite system~\cite{PuigVonFriesen2010}. Although such artificial processes would cancel in an exact theory, this is not necessarily the case when only diagrams of particular topology are summed to infinite order~\cite{PuigVonFriesen2010}. However, in unconstrained bosonic systems, the equivalent high-order processes are within the infinite-dimensional Hilbert space. In such a case, reasons for the failure of SKFT+2PI for closed systems are less clear, although it has been conjectured to also be due to neglected diagrams~\cite{Rey2004}. For constrained bosonic systems, such as the Schwinger bosons we employ to map spin operators onto, we believe that neglected non-Gaussian correlations of the initial state significantly contribute to the discrepancy between SKFT+2PI and the benchmarks in Fig.~\ref{fig:closed}, as discussed in Sec.~\ref{sec:initial}. The artifactual stationary state produced by SKFT+2PI in Fig.~\ref{fig:closed} is determined by the total energy and magnetization of the initial state, since these are conserved by the symmetries of the action in Eq.~\eqref{eq:finalAction}. Thus, in the FM system with all-but-one spins initially pointing up [illustrated in the inset of Fig.~\ref{fig:closed}(a)], the SKFT+2PI-obtained stationary state reflects these conservation laws as it converges to uniform but not saturated magnetization $\braket{\hat \sigma^z_n(t\rightarrow \infty)} < 1$. Similarly, in the AF system initially [illustrated in the inset of Fig.~\ref{fig:closed}(b)] in the \textit{unentangled} Ne\'el state, a nonzero initial magnetization (due to an odd number of spins) is conserved into the stationary limit.

\section{Conclusions and outlook}\label{sec:conclusions}


In conclusion, we construct a promising field-theoretic approach to driven-dissipative many-body systems, as one of the most challenging unsolved problems in quantum physics~\cite{DelRe2020}. This approach can tackle various dynamical regimes and system geometries or dimensionalities~\cite{Rosenberg2017,Yuan2022}. In contrast, previously developed methods, when applied to the spin-boson model as a standard testbed~\cite{Vilkoviskiy2024}, require changing the method (such as HEOM~\cite{Tanimura2020} vs. different flavors of TN approaches~\cite{Fux2024, Cygorek2024a}) depending on the chosen parameters of the model. Furthermore, we also demonstrate how our SKFT+2PI framework can handle \textit{many} interacting quantum spins with \textit{noncommuting} couplings to {\em multiple}~\cite{Bruognolo2014,Nemati2022,Hogg2024} baths generating the \textit{non-Markovian} regime of the dynamics of spins. The complexity of such a setup [Fig.~\ref{fig:fig0}(b)] causes impediments even for very recent TN methods developed~\cite{Fux2023} for this frontier problem due to transient entanglement barrier~\cite{Lerose2023,Rams2020,Foligno2023}, even in the presence of a dissipative environment~\cite{GarciaGaitan2024}. In addition, our SKFT+2PI framework can handle arbitrary spin value $S$ in arbitrary geometry~\cite{Patra2025} or spatial dimensionality~\cite{Rosenberg2017,Yuan2022}, which benefits quantum spintronics~\cite{Petrovic2021a,Suresh2024,Kovarik2024,Choi2019,Chen2023} and quantum magnonics~\cite{Yuan2022}, where the spin value $S \ge 1/2$; as well as quantum computing~\cite{Rosenberg2017} where $S=1/2$ for qubits~\cite{Gulacsi2023,Rossini2023} and $S>1/2$ for qudits~\cite{Nikolaeva2024}. 

The ability of our SKFT+2PI framework to closely track numerical (quasi)exact results from TN methods applied to the spin-boson model demonstrates that our field-theoretic approach is {\em non-perturbative} in the system-bath coupling. This achievement can be traced back to the usage of 2PI resummation of a class of infinitely many Feynman diagrams generated by the $1/N$ expansion, instead of conventional perturbative expansion in the system-bath coupling. Nevertheless, full understanding of how our 2PI and $1/N$ resum conventional perturbative expansion in the system-bath coupling constant to achieve non-perturbative regimes is lacking. It could be further clarified by constructing the associated resurgent transseries in the coupling constant~\cite{DiPietro2021} and comparing it with our expansion. In addition, our SKFT+2PI offers a \textit{single unified framework} that can handle arbitrary temperature, cutoff frequency, and spectral content of the bosonic baths (i.e., Ohmic vs. sub-Ohmic vs. super-Ohmic), or exchange interaction between many spins. Unlike widely used QMEs for open quantum system dynamics~\cite{Breuer2007}, where only the reduced density operator is accessible as a function of single time, \textit{both} our SKFT+2PI framework and TN methods~\cite{Fux2024, Cygorek2024a} make possible computation of multitime two-spin correlators. For example, using such correlators, we can obtain the order parameter [Fig.~\ref{fig:qpt}] of delocalized-localized quantum phase transition in the spin-boson model or nonequilibrium spin current [Fig.~\ref{fig:transport}] in the spin-chain-boson model biased by a temperature gradient. This can be of great interest to modeling experiments on quantum magnets, such as quantum spin liquids~\cite{Savary2017}, where thermal transport is one of the major tools~\cite{Kasahara2018}.


The numerical cost of solving integro-differential Eqs.~\eqref{eq:eomReal}(a)--\eqref{eq:eomReal}(n) produced by SKFT+2PI formalism for open quantum spin systems is cubic-scaling in the number of spins $N_S$ of arbitrary value $S$, as well as quartic-scaling in the number of time steps $N_t$. The latter computational complexity could, in principle, be lowered~\cite{Meirinhos2022, Blommel2024, Kaye2021, Zhu2025} by optimizing numerics. One of the main avenues for lowering $p$ in the scaling $\propto N_t^p$ is by introducing a memory cut, which consists of neglecting contributions to the dynamics from states before a fixed cutoff time. However, SEs and kernels related to the environment surrounding open systems may decay algebraically in time~\cite{Chakraborty2018}, rather than exponentially as in closed systems, for which memory cuts have already been implemented~\cite{Schuckert2018, Meirinhos2022, Blommel2024}. This is precisely the case for a bosonic bath with any $s$ parameter in Eq.~\eqref{eq:spectral}. As such, implementing memory cuts in open quantum systems requires additional efforts beyond those of Refs.~\cite{Schuckert2018, Meirinhos2022, Blommel2024}.

Finally, we point out that the occasional discrepancies between the SKFT+2PI framework and numerically (quasi)exact benchmarks, particularly conspicuous in Fig.~\ref{fig:strong} for ultrastrong system-bath coupling, largely stem from the truncation [Eq.~\eqref{eq:gamma2}] of the diagrammatic series, which may neglect high-energy excitations~\cite{Rey2004} or include unphysical virtual processes~\cite{PuigVonFriesen2010}. Our benchmarking efforts suggest that neglected initial higher-order correlations~\cite{Koksma2011} may also significantly contribute to such discrepancies, but this issue is much less explored in the literature. This is because non-Gaussian initial correlations are required to constrain the dynamics of the infinite-dimensional Schwinger bosons to the finite physical subspace. Including the environment in our formulation of SKFT+2PI for \textit{open} quantum spin systems alleviates the issues posed by the neglected non-Gaussian correlations by effectively erasing the initial conditions at longer times. However, when the system transitions into a localized phase, memory of the initial conditions is retained well into the stationary regime, so  deviations from numerically (quasi)exact benchmarks emerge again [Figs.~\ref{fig:strong}(a) and~\ref{fig:strong}(c)]. Therefore, finding a general and practical solution to the fundamental problem~\cite{Garny2009} of how to include \textit{arbitrary} non-Gaussian initial states into SKFT could further improve the capabilities of our approach.

\acknowledgments
We are grateful to L. E. Herrera Rodr\'iguez for technical help regarding the intricacies of HEOM calculations. F.R.-O., F.G.-G. and B.K.N. were supported by the U.S. National Science Foundation (NSF) under Grant No. DMR-2500816.  P.P. was partially supported by the U.S. Army Research Office under Grant No. W911NF-22-2-0234. S.R.C. gratefully acknowledges financial support from the UK's Engineering and Physical Sciences Research Council (EPSRC) under Grant No. EP/T028424/1.

\bibliography{references}

\end{document}